\pdfoutput=1
\documentclass[superscriptaddress,reprint,amsfonts,amsmath,amssymb,prx,nofootinbib]{revtex4-2}
\usepackage{graphicx}
\usepackage{dcolumn}
\usepackage{bm}
\usepackage[%
 setpagesize=false,%
 colorlinks=true,%
]{hyperref}
\usepackage{physics}
\usepackage{siunitx}
\usepackage{color}
\usepackage{dsfont}

\begin{document}

\title{Qutrit state discrimination with mid-circuit measurements}

\author{Naoki Kanazawa$^\dagger$}
\email[Corresponding author: ]{knzwnao@jp.ibm.com}
\affiliation{IBM Quantum, IBM Research Tokyo, Chuo-ku, Tokyo 103-8510, Japan}

\author{Haruki Emori$^\dagger$}
\affiliation{IBM Quantum, IBM Research Tokyo, Chuo-ku, Tokyo 103-8510, Japan}
\affiliation{
Graduate School of Information Science and Technology, Hokkaido University, Sapporo, Hokkaido 060-0814, Japan
}

\author{David C.~McKay}
\affiliation{IBM Quantum, IBM T.~J.~Watson Research Center, Yorktown Heights, New York 10598, USA}

\date{\today}

\begin{abstract}
Qutrit state readout is an important technology not only for execution of qutrit algorithms but also for erasure detection in error correction circuits and leakage error characterization of the gate set. 
Conventional technique using a specialized IQ discriminator requires memory intensive IQ data for input, and has difficulty in scaling up the system size.
In this study, we propose the mid-circuit measurement based discrimination technique which exploits a binary discriminator for qubit readout.
Our discriminator shows comparable performance with the IQ discriminator, and readily available for standard quantum processors calibrated for qubit control.
We also demonstrate our technique can reimplement typical benchmarking and characterization experiments such as leakage randomized benchmarking and state population decay measurement.

\end{abstract}

\maketitle

\def\thefootnote{$\dagger$}\footnotetext{These two authors contributed equally}\def\thefootnote{\arabic{footnote}}

\section{\label{sec:intro}Introduction}
Quantum computing is a rapidly evolving technology with a potential to revolutionize computation in variety of fields. 
Continuous improvements in underlying hardware and error mitigation techniques allow quantum computers to become a utility tool rather than an experimental equipment \cite{Kim_2023}. 

Among various physical systems showing quantum properties, superconducting transmon has gained significant attention.
Despite the scalability and maturity of fabrication technology, superconducting transmon is inherently an infinite energy level system with weak anharmonicity, leading to frequency collisions between neighboring energy levels \cite{Hertzberg_2021, Heya_2023}. 
This may cause a leakage outside the computational subspace. 
The derivative removal by adiabatic gate (DRAG) is a popular pulse shaping technique to mitigate such unwanted effect during the implementation of single-qubit gate \cite{Theis_2018}.
The DRAG technique allows for the gate speed-up while suppressing the leakage, and it has been extensively studied and widely used in the superconducting quantum computers \cite{Motzoi_2009, Chow_2010, Gambetta_2011, Chen_2016, Haupt_2023}.
Leakage suppression in the near adiabatic limit is also an important challenge in quantum optimal control \cite{Werninghaus_2021}.
In the context of quantum error correction, this error is also known as erasures, and undetected leakage may cause detrimental effect on the behavior of correction codes \cite{Varbanov_2020, McEwen_2021}.
As such, information of $\ket{2}$ state population is of crucial importance for gate calibration and error correction even though the quantum algorithms are usually designed for the computational basis (i.e., $\ket{0}$ and $\ket{1}$ states).

Going beyond, the use of higher energy levels offers a larger state space allowing more capacity for information and hence more efficient algorithms \cite{Roy_2022}. 
The intermediate use of a qutrit state enables efficient decomposition of multi-controlled gates such as Toffoli and Fredkin gate \cite{Gokhale_2019, Galda_2021, Saha_2023}. 
Since the dispersive readout is a standard used technique in the superconducting quantum computers, an IQ discriminator specially calibrated for the classification of ternary signals is the most commonly used approach for qutrit state readout \cite{Kurpiers_2018, Blok_2021, Chen_2023, Kehrer_2023}. 
However, such ternary classification task requires memory consuming shot-wise IQ data, and thus difficult to scale up the system size and requires the ability to calibrate the $\ket{2}$ state.
The binary outcome measurement can alleviate this overhead by reusing the readout instruction calibrated for the computational basis. 
Because $\ket{2}$ state is misclassified as $\ket{1}$ state in a typical setup, one can indirectly measure the target state by the binary measurement following an $X_{01}$ pulse which only flips the population of the computational state \cite{Cervera-Lierta_2022}.
Although this technique is useful to measure a particular state population, a qutrit state cannot be fully resolved by single trial.
If we can prepare an identical state multiple times, we can reconstruct complete probability distribution by combining a pair of measurements with and without the $X_{01}$ pulse \cite{Andrews_2019}.
Because this requires two independent state preparations, a qutrit state cannot be discriminated at one shot.

In this work, we demonstrate a technique that combines the binary outcome measurement with mid-circuit measurement (MCM) \cite{Graham_2023, Govia_2022}.
Thanks to the quantum non-demolition property, two measurement operations can be implemented back-to-back with an interleaved $X_{01}$ pulse, instead of combining two independent measurement outcomes in post-processing.
This technique allows for resolving a complete qutrit state probabilities with a single circuit, without the need for calibration of specialized IQ discriminator and measurement stimulus.
It is worth noting that the MCM-based discrimination (MCMD) technique shows a better statistical property, due to a deterministic property of the second measurement.
This is because the state to measure is projected by the first measurement.
We also experimentally confirm our technique is comparable or slightly outperforms a calibrated IQ discriminator in accuracy.

The rest of this paper is organized as follows.
In Sec.~\ref{sec:analysis}, we introduce the principle of MCMD readout along with its statistical property. 
In Sec.~\ref{sec:experiments}, we experimentally investigate how the readout error mitigation technique works with MCMD.
Multiple qubits from two quantum processors are tested to quantitatively compare its performance with the specialized qutrit IQ discriminator’s one.
Lastly, we apply the MCMD technique to typical benchmarking and characterization experiments in Sec.~\ref{sec:demo} and conclude with Sec.~\ref{sec:conclusion}.
All experiments in this study are conducted with Qiskit Experiments and IBM Quantum processors \cite{Kanazawa_2023}.

\section{\label{sec:analysis}Principle}
\begin{figure}[t]
    \includegraphics[width=0.9\linewidth]{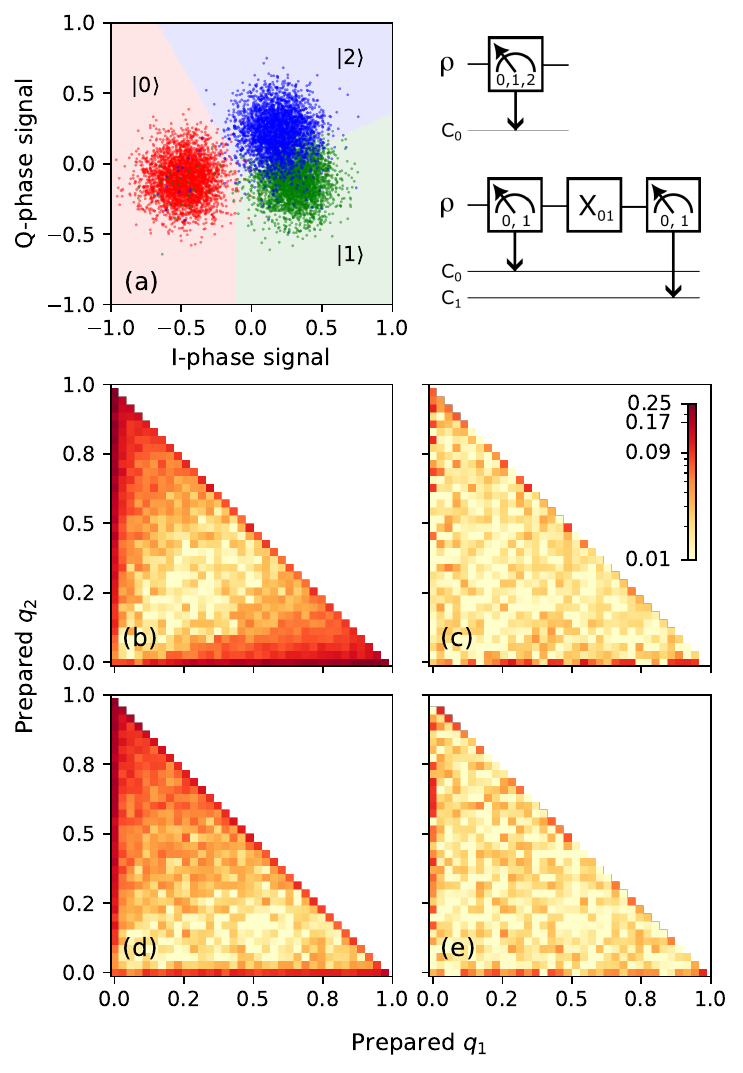}
    \caption{
        Qutrit signal discrimination. (a) Left panel: scatter plot of single-shot readout records 
        in the IQ plane. 3000 shots per each qutrit state. 
        Filled area indicates the trained IQ discrimination domain.
        Right panel: circuit diagram for qutrit state discrimination
        with a specialized IQ discriminator (top) and MCM-based discrimination circuit (bottom). 
        (b) and (d) show the distribution of Hellinger distance from the prepared state probabilities 
        for the IQ discrimination and MCM-based discrimination, respectively. 
        (c) and (e) are those with the readout error mitigation.
        See main text for experiment details.
    }
    \label{fig:probability_heatmap}
\end{figure}

Circuit diagram for the MCMD sequence is shown in Fig.~\ref{fig:probability_heatmap}. 
Suppose a binary outcome measurement $M$ with two POVM elements $M_0 = \ket{0}\!\bra{0}$ and $M_1 = \mathds{1} - \ket{0}\!\bra{0}$. 
We denote a quantum state to be measured by $\rho = \sum_{i=0}^2 a_i^2 \ket{i}\!\bra{i}$, where $\{a_i\}$ satisfy $\sum a_i^2 = 1$.
After the first measurement, the state is projected onto one of the qutrit bases $\rho_\alpha = \ket{\alpha}\!\bra{\alpha}$ and we obtain an outcome $\alpha$ at a probability $p_\alpha = \Tr[M_\alpha \rho M_\alpha^\dagger]$.
Likewise, in the second measurement we obtain an outcome $\beta$ with a probability $p_{\beta|\alpha} = \Tr[M_{\beta} X_{01} \rho_\alpha X_{01}^\dagger M_{\beta}^\dagger]$, where $X_{01} = \ket{0}\!\bra{1} + \ket{1}\!\bra{0} + e^{i\phi} \ket{2}\!\bra{2}$ and $\phi$ is imparted by the geometric and AC Stark phase \cite{Blok_2021}.
Because the second measurement is deterministic owing to the first projective measurement, we write the probability of measuring a bit sequence $\beta\alpha$ by conditional probability
\begin{equation*}
    p_{\beta\alpha} = p_{\beta|\alpha} p_\alpha = \Tr[M_\beta X_{01} M_\alpha \rho M_\alpha^\dagger X_{01}^\dagger M_\beta^\dagger].
\end{equation*}
This gives four probabilities $p_{00} = 0$, $p_{10} = a_0^2$, $p_{01} = a_1^2$, and $p_{11} = a_2^2$.
Each two bit outcome is straightforwardly associated with the corresponding qutrit state, except for $00$ which may appear due to some imperfection.

Next, we consider the non-MCM case used in Ref.\cite{Andrews_2019}, in which we prepare a reproducible state $\rho$ and measure it in two circuits with and without the $X_{01}$ pulse.
Although this protocol is similar with MCMD, two measurements are statistically independent and probabilities for two outcomes are written as follows.
\begin{eqnarray*}
    p_\alpha &=& \Tr[M_{\alpha} \rho M_{\alpha}^\dagger], \\
    p_\beta &=& \Tr[M_{\beta} X_{01} \rho X_{01}^\dagger M_{\beta}^\dagger].
\end{eqnarray*}
This yields $p_{\alpha, 0} = a_0^2$ and $p_{\alpha, 1} = a_1^2 + a_2^2$ for an outcome $\alpha$, and $p_{\beta, 0} = a_1^2$ and $p_{\beta, 1} = a_0^2 + a_2^2$ for an outcome $\beta$.
Solving these outcomes for the $\ket{2}$ state probability yields $a_2^2 = p_{\alpha, 1} + p_{\beta, 1} - 1$.

We call attention that the probability of measuring some outcome $x$ with a finite number of trial $N$ results in a finite variance arising from the sampling error, which might be written as $V(x) = N p_x (1 - p_x)$ by assuming the binomial distribution. 
When we measure a $\ket{2}$ state probability of some quantum circuit with MCMD, the variance is estimated to be $V_{\text MCM} = V(p_{11}) = N a_2^2 (1 - a_2^2)$.
On the other hand, that of the independent measurement becomes $V_{\text IM} = V(p_{\alpha, 1}) + V(p_{\beta, 1}) = N\left( a_0^2 (1 - a_0^2) + a_1^2 (1 - a_1^2) \right)$.
This implies MCMD shows smaller variance
\begin{equation*}
    \frac{V_{\text MCM}}{V_{\text IM}} = 1 - \frac{2 a_0^2 a_1^2}{a_0^2 (1 - a_0^2) + a_1^2 (1 - a_1^2)} \leq 1
\end{equation*}
regardless of the measured state, because $0 \leq a_i \leq 1$.
This is beneficial especially when we measure a small $\ket{2}$ probability, such as characterizing a leakage error of the calibrated gate set.

\section{\label{sec:experiments}Experiments}
A potential side effect of MCMD is unwanted state transition during the first measurement \cite{Sank_2016}.
Because the state after measurement immediately experiences the $X_{01}$ pulse and the second measurement, the dynamics of the measurement error can be quite complex.
We experimentally investigate how the conventional readout error mitigation (REM) technique \cite{Maciejewski_2020} improves MCMD outcomes.

We run experiments on two superconducting quantum processors IBM Quantum provides; \texttt{ibm\_canberra} (Falcon R6) and \texttt{ibm\_algiers} (Falcon R5.11) with the measurement instruction duration of $1632$ ns and $857$ ns, respectively.
As a reference, we calibrate an IQ discriminator specialized for the qutrit readout. 
In this study, we exploit a classifier with a quadratic decision boundary implemented by the scikit-learn library \cite{scikit-learn}.
We expect that the variance in the ideal IQ discriminator measurement is comparable with MCMD, because it can also fully resolve the qutrit probability distribution with a single circuit.
We prepare an arbitrary qutrit probability distribution with a custom calibrated pulse $X_{12} = \ket{1}\!\bra{2} + \ket{2}\!\bra{1} + e^{i\phi} \ket{0}\!\bra{0}$ along with the SX and virtual Z gate \cite{McKay_2017} provided by IBM processors.
For MCMD readout, we delegate the IQ discrimination task to the IBM hardware discriminator that only returns a count for the binary outcome.
Details concerning the calibration and state preparation circuits are described in Appendix.~\ref{sec:app.calibration}.
A typical experiment result from the qubit $0$ of \texttt{ibm\_canberra} is shown in Fig.~\ref{fig:probability_heatmap}.
Heat maps in the panel (b) and (d) show the unmitigated error of the specialized IQ discriminator and MCMD, respectively.
Each data point is sampled $1000$ times, and the measured probability distribution $u = (p_0, p_1, p_2)$ is compared with the prepared distribution $v = (q_0, q_1, q_2)$.
The error from the prepared distribution is measured by the Hellinger distance
\begin{equation*}
    d_H(u, v) = \left( \sum_{i=0}^2 (\sqrt{p_i} - \sqrt{q_i})^2 \right)^{1/2},
\end{equation*}
where $0 \leq d_H \leq \sqrt{2}$.
The Hellinger distance increases as $q_1$ and $q_2$ increase in the IQ discriminator, resulting from the significant overlap of the kerneled signal of the $\ket{1}$ and $\ket{2}$ state in the IQ plane, as shown in Fig.~\ref{fig:probability_heatmap}(a).
This is a common behavior in a quantum processor with dispersive readout stimuli which are not calibrated for qutrit signals.
This overlap becomes an obstacle to implement an accurate IQ discriminator without REM.
Contrariwise, the increase of the Hellinger distance in MCMD is only visible in large $q_2$.
This implies the confusion of $\ket{0}$ and $\ket{2}$ signal in the IBM hardware discriminator or relaxation of $\ket{2}$ into $\ket{1}$ state as we concern.
However, MCMD is not sensitive to the overlap of $\ket{1}$ and $\ket{2}$ signal and shows smaller overall error distance without REM.

To apply REM, we first need to experimentally measure an invertible map $\Lambda_{ij}$ that represents the effect of classical noise on POVM $M^{\rm exp}_i = \Lambda_{ij} M_j$ where $M_j = \ket{j}\!\bra{j}$.
Given the noise is invertible, we can write quasi probability $v = \Lambda^{-1} u$,
which is mapped to the closest probability defined by L2 norm \cite{Smolin_2012}.
It should also be borne in mind that we often observe a finite immaterial $p_{00}$ probability in MCMD, resulting in $i \in [00, 01, 10, 11]$ and $j \in [0, 1, 2]$. 
In other words, the inverse map $\Lambda^{-1}$ is not a square matrix and it is computed by pseudo inverse.
Although this is a special case in REM, the mean $p_{00}$ value is usually small; $4.81 \times 10^{-3}$ and $9.84 \times 10^{-3}$ in \texttt{ibm\_canberra} and \texttt{ibm\_algiers}, respectively.
In both processors, $p_{00}$ value has a tendency to be large when a $\ket{0}$ state is measured, indicating the heating during the first measurement, or the error of the interleaved $X_{01}$ gate due to residual cavity photons \cite{McClure_2016, rudinger2021characterizing}.
Investigation of the detailed error mechanism is out of scope in this study.
Nevertheless, the conventional REM technique works fine in both IQ discriminator and MCMD,
as shown in Fig.~\ref{fig:probability_heatmap} (c) and (e).

\begin{figure}[t]
    \includegraphics[width=1.0\linewidth]{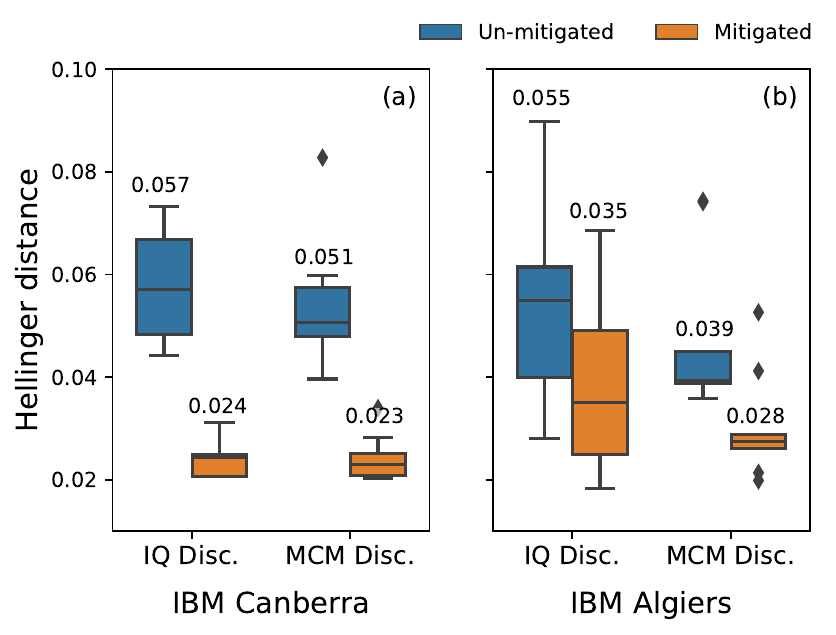}
    \caption{
        Statistics of the mean Hellinger distance sampled from representitive $9$ qubits
        in (a) \texttt{ibm\_canberra} and (b) \texttt{ibm\_algiers}.
        The number shows the $50$th percentile of the mean Hellinger distance on
        each configuration.
        See the main text for experiment details.
    }
    \label{fig:hellinger_boxplot}
\end{figure}

We repeat this experiment for multiple qubits on \texttt{ibm\_canberra} and \texttt{ibm\_algiers}, and average the Helliger distance over all prepared quantum states in each qubit.
To minimize the time window of the entire experiment, every task on non adjacent qubits is batched together and executed in parallel. 
This alleviates the device parameter drift during the experiment.
Because the specialized IQ discriminator requires a massive raw kerneled data sent over the wire, we only choose random $9$ qubits out of available $27$ qubits for the experiment.
As shown in Fig.~\ref{fig:hellinger_boxplot}, we verified the efficacy of REM on MCMD at the processor scale.
The effectiveness of REM is comparable with the IQ discriminator. 
The reduction of the error distance measured by $50$th percentile for the IQ discriminator (MCMD) experiment is $57.9$\% ($54.9$\%) in \texttt{ibm\_canberra}, and $36.3$\% ($28.2$\%) in \texttt{ibm\_algiers}.
The magnitude of the mitigated Hellinger distance is comparable or even smaller in MCMD.
Change in the Hellinger distance distribution by REM is also available in Appendix \ref{sec:app.hellinger_data} for individual qubits.

\section{\label{sec:demo}Applications}
\begin{figure}[t]
    \includegraphics[width=1.0\linewidth]{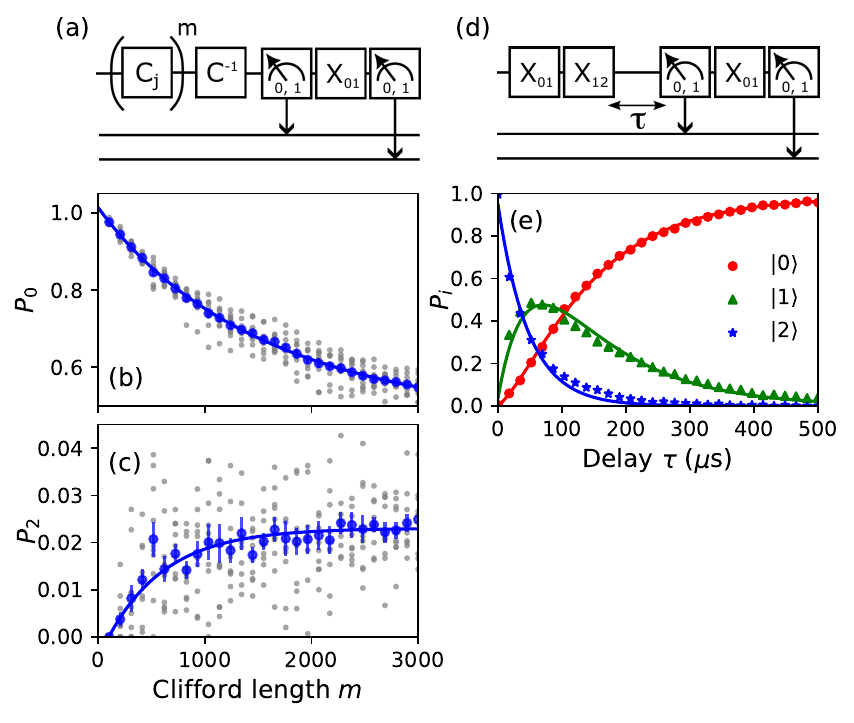}
    \caption{
        Benchmarking and characterization experiments using the MCMD technique. 
        (a) Circuit diagram of the leakage RB with MCMD. 
        $C_j$ and $C^{-1}$ are Clifford gates in the computational basis.
        Measured population of $\ket{0}$ and $\ket{2}$ states are shown in (b) and (c).
        Gray dots are outcome of each trial, and blue dots are the average of them.
        (e) Circuit diagram of state population decay measurement.
        Prepared $\ket{2}$ state is measured with MCMD after delay $\tau$. 
        Observed dynamics of each qutrit state is shown in (e).
        All probabilities are computed with REM.
    }
    \label{fig:mcmd_demo}
\end{figure}

To confirm the usefulness of our technique, we apply MCMD to benchmarking and characterization.
We first implement leakage randomized benchmarking (LRB) experiment with MCMD readout. 
This experiment is comparable with the standard Clifford RB experiment, but $\ket{2}$ state population is measured instead of the survival probability, which allows us to evaluate a leakage error of the gate set.
The experimental circuit is illustrated in Fig.~\ref{fig:mcmd_demo} (a).
A random Clifford sequence of length $m$ followed by another Clifford $C^{-1}$ inverting the previous action is constructed, and we resolve the final qutrit state with the MCMD readout.
The state population at Clifford length $m$ can be fit by
\begin{eqnarray*}
    p_2(m) = 1 - A - B \lambda ^ m,
\end{eqnarray*}
where $A, B$, and $\lambda$ are all fitting parameters.
By using the fit values, we compute 
\begin{eqnarray*}
    L_1 &=& (1-A)(1-\lambda), \\
    L_2 &=& A (1-\lambda),
\end{eqnarray*}
which we call average leakage rate and average seepage rate, respectively \cite{Wood_2018}.
The leakage rate quantifies a population to the non computational state, 
and the seepage rate quantifies a return population back into the computational state from the non computational state, which is driven by the relaxation from $\ket{2}$ state.

Because a gate set provided by an IBM processor is usually tuned to suppress leakage errors, we intentionally calibrate a custom SX gate to cause a visible leakage for this demonstration, and decompose a Clifford gate into two SX gates with virtual Z gates.
The leaky gate is implemented with a DRAG pulse with Gaussian $\sigma = 1.78$ ns and duration $t_g = 35.5$ ns, which is $5$ times faster than a typical value in IBM gate set.
Such pulse has a broader frequency spectrum and strongly drives a qubit, contributing to the leakage error.
We run this experiment on the qubit $0$ of \texttt{ibm\_canberra}.
The leaky gate is calibrated by a standard procedure for single qubit gates without calibration for the drive frequency.
$10$ random sequences are prepared for each $m$, and each circuit is sampled $3000$ times.
The result is shown in Fig.~\ref{fig:mcmd_demo} (b) and (c).
Since MCMD allows for measuring the $\ket{0}$ state population together, we can also compute the average gate error at no additional cost.
For this leaky gate, the average gate error is $3.1(5) \times 10^{-4}$, and the average leakage and seepage rate per Clifford are $L_1 = 4.2(4) \times 10^{-5}$ and  $L_2 = 1.8(2) \times 10^{-3}$, respectively.

To validate the experimental result, we separately measure the relaxation dynamics of $\ket{2}$ state by varying the time delay before the MCMD readout as shown in Fig.~\ref{fig:mcmd_demo} (d), which allows us to roughly estimate the lower bound of the seepage rate.
The relaxation dynamics of a qutrit state $p = (p_0, p_1, p_2)^\mathsf{T}$ can be described by rate equation of a three level system 
\begin{eqnarray*}
    \dv{p}{t} = \Gamma^\mathsf{T} p,
\end{eqnarray*}
where $\Gamma_{ij}$ is a transition rate from $\ket{i}$ to $\ket{j}$.
$\Gamma_{ij} = 0$ when $i < j$ because thermal excitation is ignored at the operating temperature of the processor, and $\Gamma_{jj} = - \sum_{j=0}^{j-1} \Gamma_{jk}$ \cite{Peterer_2015}.
The result is shown in Fig.~\ref{fig:mcmd_demo} (e).
$\ket{2}$ state is almost completely relaxed to ground state after $500$ $\mathrm{\mu s}$ from state preparation.
By fitting the experimental data with the rate equation, we obtain $\Gamma_{10}^{-1} = 118$ $\mathrm{\mu s}$, $\Gamma_{20}^{-1} = 417$ $\mathrm{\mu s}$, and $\Gamma_{21}^{-1} = 53.4$ $\mathrm{\mu s}$.
The population relaxed from $\ket{2}$ state during a single Clifford gate is $2 t_g (\Gamma_{20} + \Gamma_{21}) = 1.5 \times 10^{-3}$, which is pretty consistent with the LRB experiment.
Actual seepage rate becomes slightly higher because a quantum state is more susceptible to noise during the microwave drive.
This result indicates MCMD measures the non computational state population quite accurately, regardless of the experimental circuits.

\section{\label{sec:conclusion}Conclusion}
In this work we demonstrated a use of mid-circuit measurement combined with a binary POVM to resolve a full qutrit state probability.
The readout fidelity of our MCMD technique is comparable or even slightly better compared with the conventional approach of using a specialized qutrit IQ discriminator.

IQ discriminator classifies an IQ signal plane into multiple subdomains corresponding to the measured states.
In contrast to the qutrit discriminator in which we require memory consuming input data and computationally expensive mathematical functions, an IQ discriminator for qubit readout just requires a threshold voltage along the principal readout axis and thus hardware efficient \cite{Bronn_2017}.
Qubit discriminator can be easily realized as a hardware-implemented subroutine, providing a faster feedback loop necessary for conditional quantum operations \cite{Fu_2018}.
MCMD can reuse such hardware native instruction.
This drastically reduces the computational overhead for qutrit readout, and MCMD may enable for erasure detection in QEC circuit and execution of qutrit algorithms also in a control system with severe enegy constratins, such as in cryo-CMOS controllers \cite{Underwood_2023}.

It is noteworthy that MCMD is calibration-free, indicating we can access the second excitation state information of commercial quantum processors which may not provide a dedicated instruction for qutrit readout.
This may allow for further study on the qubit layout optimization, considering the leakage error of the underlying qubits.

\acknowledgements
N.K. and H.E. thank Itoko Toshinari for advice on readout error mitigation for MCMD readout.

\appendix

\section{\label{sec:app.calibration}Calibration of state preparation circuit}
\begin{figure*}[t!]
    \centering
    \includegraphics[width=1.3\columnwidth]{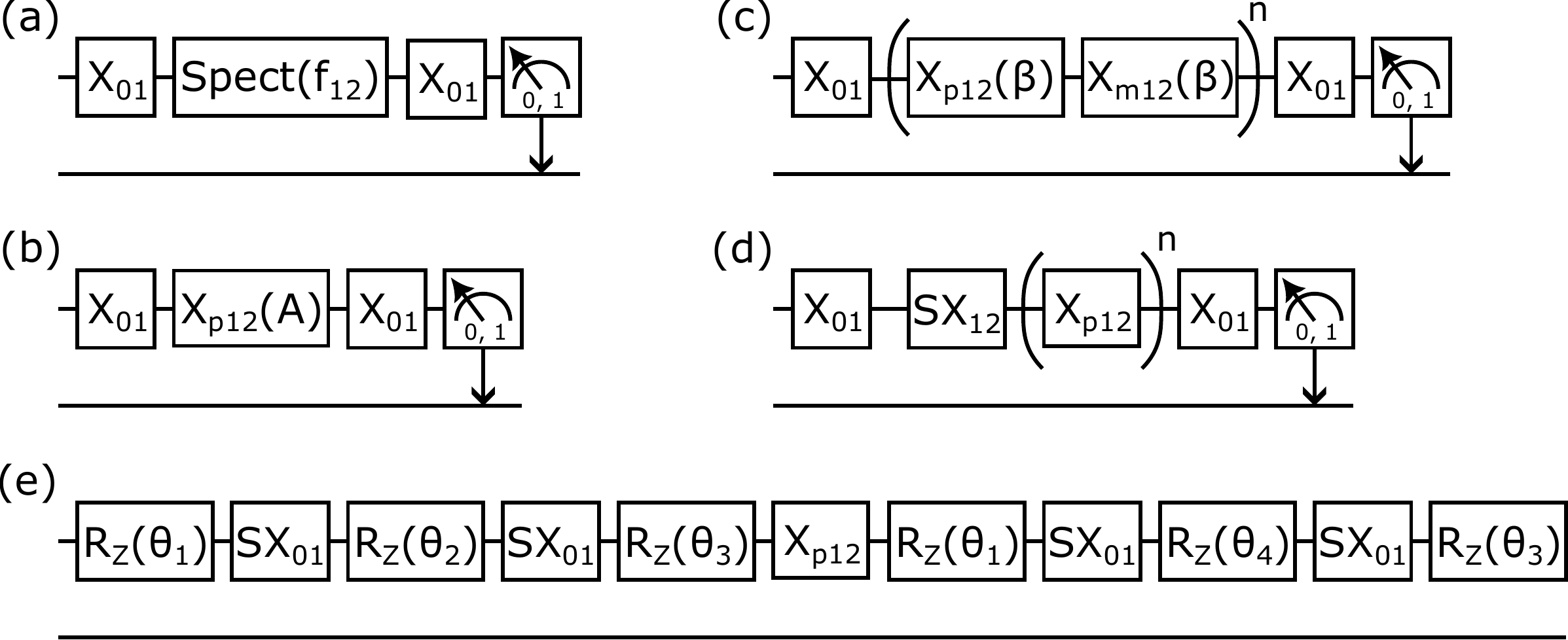}
    \caption{
        Experimental circuits used for calibration and qutrit state preparation.
        (a) Frequency calibration, 
        (b) Amplitude calibration, 
        (c) DRAG parameter calibration, 
        (d) Amplitude error amplification,
        and (e) Qutrit state preparation circuit.
        Subscript numbers denote a subspace that the gate acts on, and p ($0$) and m ($\pi$) indicates the phase of the pulse. ${\rm X}$ and ${\rm SX}$ are the $\pi$ and half-$\pi$ rotation about X axis.
        ${\rm R_Z}(\theta_i)$ is the Z rotation at angle $\theta_i$ in the computational (01) subspace implemented by the virtual Z gate.
        Instructions in the 01 subspace are all provided by the IBM processors.
    }
    \label{fig:appendix_circuits}
\end{figure*}

We calibrate the ${\rm X}_{12}$ gate which is required for qutrit state preparation with operations shown in Fig.~\ref{fig:appendix_circuits}(a) -- (d).
The ${\rm X}_{12}$ gate is implemented by a DRAG pulse with three parameters to calibrate; 
pulse amplitude $A$, DRAG parameter $\beta$, and drive frequency $f_{12}$.
Pulse duration and Gaussian $\sigma$ are fixed to $35.5$ ns and $8.89$ ns, respectively, which are the typical values in the IBM gate set for the computational state control.

We begin with the rough frequency calibration shown in Fig.~\ref{fig:appendix_circuits}(a), in which the Spect gate consists of a Gaussian pulse with $\sigma = 62.2$ ns.
This is followed by the rough amplitude calibration in Fig.~\ref{fig:appendix_circuits}(b), which scans the pulse amplitude to find a rough estimate of amplitude for $\pi$ and half-$\pi$ rotation.
The calibrated $\pi$ amplitude $A_{\pi}$ is used for the fine frequency calibration with a different Gaussian pulse with a larger $\sigma' = \Delta f^{-1}$, where $\Delta f = 1$ MHz is the target frequency resolution.
The amplitude of the spectroscopy pulse is determined by $A_{\pi} \sigma / \sigma'$,
which roughly implements the $\pi$ rotation to maximize the contrast of the measured spectrum.
Such narrow band scan results in double peak spectrum because of the charge dispersion in $\ket{2}$ state \cite{Blok_2021}, and we fit the measured spectrum with two Lorentz functions and take a middle frequency for the qutrit drive.
Next, we scan the DRAG parameter $\beta$ with the circuit in Fig.~\ref{fig:appendix_circuits}(c) with $n = 1, 3, 5$.
The optimal $\beta$ is found at the common minimum $p_1$ for all $n$.
Lastly, remaining error in the amplitude $A_{\pi}$ is removed by the error amplification circuit in Fig.~\ref{fig:appendix_circuits}(d) with $n=[0, 14]$ \cite{Sheldon_2016}.
We omit the fine calibration for the DRAG parameter $\beta$ because qutrit Stark shift is unstable due to the background charge fluctuation.
Also, the Stark shift in the $01$ subspace is not calibrated for the same reason.

This protocol gives us a sufficiently reliable $X_{12}$ gate for state preparation shown in Fig.~\ref{fig:appendix_circuits}(e). 
We control the qutrit state probability distribution $(p_0, p_1, p_2)$ with four Z rotation parameters
\begin{eqnarray*}
    \theta_1 &=& \frac{\pi}{2}, \\
    \theta_2 &=& 2 \cos^{-1}\left(\sqrt{p_0 + p_1}\right) + \pi, \\
    \theta_3 &=& \frac{5}{2} \pi, \\
    \theta_4 &=& 2 \cos^{-1}\left(\sqrt{\frac{p_0}{p_0 + p_1}}\right) + \pi,
\end{eqnarray*}
where $\forall p_i \geq 0$ and $\sum p_i = 1$, and $\theta_4 = \pi$ when $p_2 = 1$.
Geometric (Berry) phase associated with the X rotations is considered in the Z rotations.

\section{\label{sec:app.hellinger_data}Full Hellinger distance data}
\begin{figure*}[t!]
    \centering
    \includegraphics[width=2.0\columnwidth]{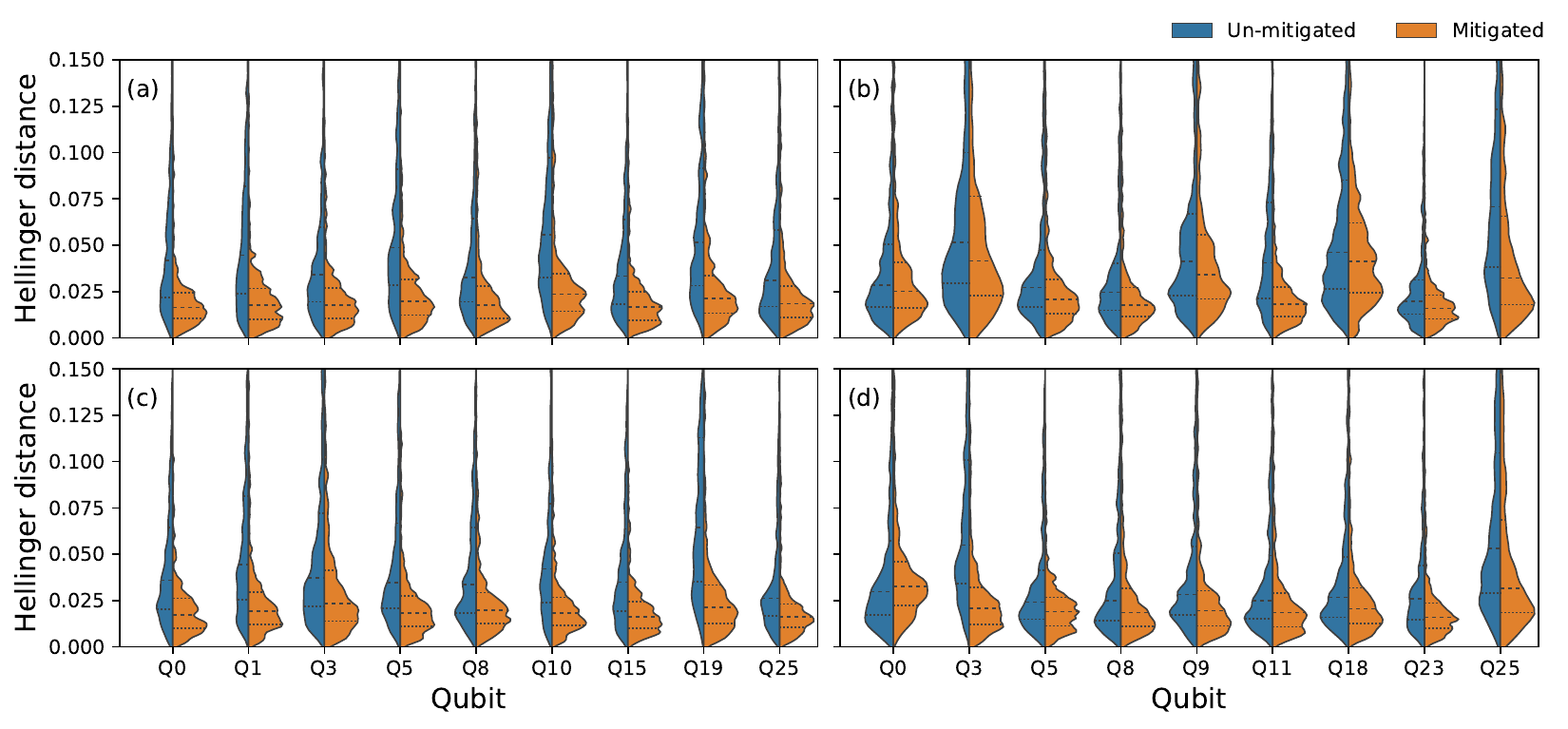}
    \caption{
        Change in the Hellinger distance distribution of every qubit in the error mitigation experiment (Sec.~\ref{sec:experiments}).
        (a, b) are the results of the specialized IQ discriminator in \texttt{ibm\_canberra} and \texttt{ibm\_algiers}, respectively.
        (c, d) are the results of the MCMD readout for the same device.
    }
    \label{fig:hellinger_viorins}
\end{figure*}

A full set of Hellinger distance data is shown in Fig.~\ref{fig:hellinger_viorins}.
These violin plots show the change in distribution of the Hellinger distance before and after application of the REM for all measured qubits.
The $50$th percentile value is reduced after application of the REM, except for the MCMD on qubit $0$ of \texttt{ibm\_algiers}.

\clearpage

\bibliography{references}

\begin{thebibliography}{38}%
\makeatletter
\providecommand \@ifxundefined [1]{%
 \@ifx{#1\undefined}
}%
\providecommand \@ifnum [1]{%
 \ifnum #1\expandafter \@firstoftwo
 \else \expandafter \@secondoftwo
 \fi
}%
\providecommand \@ifx [1]{%
 \ifx #1\expandafter \@firstoftwo
 \else \expandafter \@secondoftwo
 \fi
}%
\providecommand \natexlab [1]{#1}%
\providecommand \enquote  [1]{``#1''}%
\providecommand \bibnamefont  [1]{#1}%
\providecommand \bibfnamefont [1]{#1}%
\providecommand \citenamefont [1]{#1}%
\providecommand \href@noop [0]{\@secondoftwo}%
\providecommand \href [0]{\begingroup \@sanitize@url \@href}%
\providecommand \@href[1]{\@@startlink{#1}\@@href}%
\providecommand \@@href[1]{\endgroup#1\@@endlink}%
\providecommand \@sanitize@url [0]{\catcode `\\12\catcode `\$12\catcode
  `\&12\catcode `\#12\catcode `\^12\catcode `\_12\catcode `\%12\relax}%
\providecommand \@@startlink[1]{}%
\providecommand \@@endlink[0]{}%
\providecommand \url  [0]{\begingroup\@sanitize@url \@url }%
\providecommand \@url [1]{\endgroup\@href {#1}{\urlprefix }}%
\providecommand \urlprefix  [0]{URL }%
\providecommand \Eprint [0]{\href }%
\providecommand \doibase [0]{https://doi.org/}%
\providecommand \selectlanguage [0]{\@gobble}%
\providecommand \bibinfo  [0]{\@secondoftwo}%
\providecommand \bibfield  [0]{\@secondoftwo}%
\providecommand \translation [1]{[#1]}%
\providecommand \BibitemOpen [0]{}%
\providecommand \bibitemStop [0]{}%
\providecommand \bibitemNoStop [0]{.\EOS\space}%
\providecommand \EOS [0]{\spacefactor3000\relax}%
\providecommand \BibitemShut  [1]{\csname bibitem#1\endcsname}%
\let\auto@bib@innerbib\@empty
\bibitem [{\citenamefont {Kim}\ \emph {et~al.}(2023)\citenamefont {Kim},
  \citenamefont {Eddins}, \citenamefont {Anand}, \citenamefont {Wei},
  \citenamefont {van~den Berg}, \citenamefont {Rosenblatt}, \citenamefont
  {Nayfeh}, \citenamefont {Wu}, \citenamefont {Zaletel}, \citenamefont
  {Temme},\ and\ \citenamefont {Kandala}}]{Kim_2023}%
  \BibitemOpen
  \bibfield  {author} {\bibinfo {author} {\bibfnamefont {Y.}~\bibnamefont
  {Kim}}, \bibinfo {author} {\bibfnamefont {A.}~\bibnamefont {Eddins}},
  \bibinfo {author} {\bibfnamefont {S.}~\bibnamefont {Anand}}, \bibinfo
  {author} {\bibfnamefont {K.~X.}\ \bibnamefont {Wei}}, \bibinfo {author}
  {\bibfnamefont {E.}~\bibnamefont {van~den Berg}}, \bibinfo {author}
  {\bibfnamefont {S.}~\bibnamefont {Rosenblatt}}, \bibinfo {author}
  {\bibfnamefont {H.}~\bibnamefont {Nayfeh}}, \bibinfo {author} {\bibfnamefont
  {Y.}~\bibnamefont {Wu}}, \bibinfo {author} {\bibfnamefont {M.}~\bibnamefont
  {Zaletel}}, \bibinfo {author} {\bibfnamefont {K.}~\bibnamefont {Temme}},\
  and\ \bibinfo {author} {\bibfnamefont {A.}~\bibnamefont {Kandala}},\ }\href
  {https://doi.org/10.1038/s41586-023-06096-3} {\bibfield  {journal} {\bibinfo
  {journal} {Nature}\ }\textbf {\bibinfo {volume} {618}},\ \bibinfo {pages}
  {500} (\bibinfo {year} {2023})}\BibitemShut {NoStop}%
\bibitem [{\citenamefont {Hertzberg}\ \emph {et~al.}(2021)\citenamefont
  {Hertzberg}, \citenamefont {Zhang}, \citenamefont {Rosenblatt}, \citenamefont
  {Magesan}, \citenamefont {Smolin}, \citenamefont {Yau}, \citenamefont
  {Adiga}, \citenamefont {Sandberg}, \citenamefont {Brink}, \citenamefont
  {Chow},\ and\ \citenamefont {Orcutt}}]{Hertzberg_2021}%
  \BibitemOpen
  \bibfield  {author} {\bibinfo {author} {\bibfnamefont {J.~B.}\ \bibnamefont
  {Hertzberg}}, \bibinfo {author} {\bibfnamefont {E.~J.}\ \bibnamefont
  {Zhang}}, \bibinfo {author} {\bibfnamefont {S.}~\bibnamefont {Rosenblatt}},
  \bibinfo {author} {\bibfnamefont {E.}~\bibnamefont {Magesan}}, \bibinfo
  {author} {\bibfnamefont {J.~A.}\ \bibnamefont {Smolin}}, \bibinfo {author}
  {\bibfnamefont {J.-B.}\ \bibnamefont {Yau}}, \bibinfo {author} {\bibfnamefont
  {V.~P.}\ \bibnamefont {Adiga}}, \bibinfo {author} {\bibfnamefont
  {M.}~\bibnamefont {Sandberg}}, \bibinfo {author} {\bibfnamefont
  {M.}~\bibnamefont {Brink}}, \bibinfo {author} {\bibfnamefont {J.~M.}\
  \bibnamefont {Chow}},\ and\ \bibinfo {author} {\bibfnamefont {J.~S.}\
  \bibnamefont {Orcutt}},\ }\href {https://doi.org/10.1038/s41534-021-00464-5}
  {\bibfield  {journal} {\bibinfo  {journal} {npj Quantum Inf.}\ }\textbf
  {\bibinfo {volume} {7}},\ \bibinfo {pages} {129} (\bibinfo {year}
  {2021})}\BibitemShut {NoStop}%
\bibitem [{\citenamefont {Heya}\ \emph {et~al.}(2023)\citenamefont {Heya},
  \citenamefont {Malekakhlagh}, \citenamefont {Merkel}, \citenamefont
  {Kanazawa},\ and\ \citenamefont {Pritchett}}]{Heya_2023}%
  \BibitemOpen
  \bibfield  {author} {\bibinfo {author} {\bibfnamefont {K.}~\bibnamefont
  {Heya}}, \bibinfo {author} {\bibfnamefont {M.}~\bibnamefont {Malekakhlagh}},
  \bibinfo {author} {\bibfnamefont {S.}~\bibnamefont {Merkel}}, \bibinfo
  {author} {\bibfnamefont {N.}~\bibnamefont {Kanazawa}},\ and\ \bibinfo
  {author} {\bibfnamefont {E.}~\bibnamefont {Pritchett}},\ }\href
  {https://doi.org/10.48550/arXiv.2302.12816} {\bibfield  {journal} {\bibinfo
  {journal} {arXiv.2302.12816}\ } (\bibinfo {year} {2023})}\BibitemShut
  {NoStop}%
\bibitem [{\citenamefont {Theis}\ \emph {et~al.}(2018)\citenamefont {Theis},
  \citenamefont {Motzoi}, \citenamefont {Machnes},\ and\ \citenamefont
  {Wilhelm}}]{Theis_2018}%
  \BibitemOpen
  \bibfield  {author} {\bibinfo {author} {\bibfnamefont {L.~S.}\ \bibnamefont
  {Theis}}, \bibinfo {author} {\bibfnamefont {F.}~\bibnamefont {Motzoi}},
  \bibinfo {author} {\bibfnamefont {S.}~\bibnamefont {Machnes}},\ and\ \bibinfo
  {author} {\bibfnamefont {F.~K.}\ \bibnamefont {Wilhelm}},\ }\href
  {https://doi.org/10.1209/0295-5075/123/60001} {\bibfield  {journal} {\bibinfo
   {journal} {Europhys. Lett.}\ }\textbf {\bibinfo {volume} {123}},\ \bibinfo
  {pages} {60001} (\bibinfo {year} {2018})}\BibitemShut {NoStop}%
\bibitem [{\citenamefont {Motzoi}\ \emph {et~al.}(2009)\citenamefont {Motzoi},
  \citenamefont {Gambetta}, \citenamefont {Rebentrost},\ and\ \citenamefont
  {Wilhelm}}]{Motzoi_2009}%
  \BibitemOpen
  \bibfield  {author} {\bibinfo {author} {\bibfnamefont {F.}~\bibnamefont
  {Motzoi}}, \bibinfo {author} {\bibfnamefont {J.~M.}\ \bibnamefont
  {Gambetta}}, \bibinfo {author} {\bibfnamefont {P.}~\bibnamefont
  {Rebentrost}},\ and\ \bibinfo {author} {\bibfnamefont {F.~K.}\ \bibnamefont
  {Wilhelm}},\ }\href {https://doi.org/10.1103/PhysRevLett.103.110501}
  {\bibfield  {journal} {\bibinfo  {journal} {Phys. Rev. Lett.}\ }\textbf
  {\bibinfo {volume} {103}},\ \bibinfo {pages} {110501} (\bibinfo {year}
  {2009})}\BibitemShut {NoStop}%
\bibitem [{\citenamefont {Chow}\ \emph {et~al.}(2010)\citenamefont {Chow},
  \citenamefont {DiCarlo}, \citenamefont {Gambetta}, \citenamefont {Motzoi},
  \citenamefont {Frunzio}, \citenamefont {Girvin},\ and\ \citenamefont
  {Schoelkopf}}]{Chow_2010}%
  \BibitemOpen
  \bibfield  {author} {\bibinfo {author} {\bibfnamefont {J.~M.}\ \bibnamefont
  {Chow}}, \bibinfo {author} {\bibfnamefont {L.}~\bibnamefont {DiCarlo}},
  \bibinfo {author} {\bibfnamefont {J.~M.}\ \bibnamefont {Gambetta}}, \bibinfo
  {author} {\bibfnamefont {F.}~\bibnamefont {Motzoi}}, \bibinfo {author}
  {\bibfnamefont {L.}~\bibnamefont {Frunzio}}, \bibinfo {author} {\bibfnamefont
  {S.~M.}\ \bibnamefont {Girvin}},\ and\ \bibinfo {author} {\bibfnamefont
  {R.~J.}\ \bibnamefont {Schoelkopf}},\ }\href
  {https://doi.org/10.1103/PhysRevA.82.040305} {\bibfield  {journal} {\bibinfo
  {journal} {Phys. Rev. A}\ }\textbf {\bibinfo {volume} {82}},\ \bibinfo
  {pages} {040305} (\bibinfo {year} {2010})}\BibitemShut {NoStop}%
\bibitem [{\citenamefont {Gambetta}\ \emph {et~al.}(2011)\citenamefont
  {Gambetta}, \citenamefont {Motzoi}, \citenamefont {Merkel},\ and\
  \citenamefont {Wilhelm}}]{Gambetta_2011}%
  \BibitemOpen
  \bibfield  {author} {\bibinfo {author} {\bibfnamefont {J.~M.}\ \bibnamefont
  {Gambetta}}, \bibinfo {author} {\bibfnamefont {F.}~\bibnamefont {Motzoi}},
  \bibinfo {author} {\bibfnamefont {S.~T.}\ \bibnamefont {Merkel}},\ and\
  \bibinfo {author} {\bibfnamefont {F.~K.}\ \bibnamefont {Wilhelm}},\ }\href
  {https://doi.org/10.1103/PhysRevA.83.012308} {\bibfield  {journal} {\bibinfo
  {journal} {Phys. Rev. A}\ }\textbf {\bibinfo {volume} {83}},\ \bibinfo
  {pages} {012308} (\bibinfo {year} {2011})}\BibitemShut {NoStop}%
\bibitem [{\citenamefont {Chen}\ \emph {et~al.}(2016)\citenamefont {Chen},
  \citenamefont {Kelly}, \citenamefont {Quintana}, \citenamefont {Barends},
  \citenamefont {Campbell}, \citenamefont {Chen}, \citenamefont {Chiaro},
  \citenamefont {Dunsworth}, \citenamefont {Fowler}, \citenamefont {Lucero},
  \citenamefont {Jeffrey}, \citenamefont {Megrant}, \citenamefont {Mutus},
  \citenamefont {Neeley}, \citenamefont {Neill}, \citenamefont {O'Malley},
  \citenamefont {Roushan}, \citenamefont {Sank}, \citenamefont {Vainsencher},
  \citenamefont {Wenner}, \citenamefont {White}, \citenamefont {Korotkov},\
  and\ \citenamefont {Martinis}}]{Chen_2016}%
  \BibitemOpen
  \bibfield  {author} {\bibinfo {author} {\bibfnamefont {Z.}~\bibnamefont
  {Chen}}, \bibinfo {author} {\bibfnamefont {J.}~\bibnamefont {Kelly}},
  \bibinfo {author} {\bibfnamefont {C.}~\bibnamefont {Quintana}}, \bibinfo
  {author} {\bibfnamefont {R.}~\bibnamefont {Barends}}, \bibinfo {author}
  {\bibfnamefont {B.}~\bibnamefont {Campbell}}, \bibinfo {author}
  {\bibfnamefont {Y.}~\bibnamefont {Chen}}, \bibinfo {author} {\bibfnamefont
  {B.}~\bibnamefont {Chiaro}}, \bibinfo {author} {\bibfnamefont
  {A.}~\bibnamefont {Dunsworth}}, \bibinfo {author} {\bibfnamefont {A.~G.}\
  \bibnamefont {Fowler}}, \bibinfo {author} {\bibfnamefont {E.}~\bibnamefont
  {Lucero}}, \bibinfo {author} {\bibfnamefont {E.}~\bibnamefont {Jeffrey}},
  \bibinfo {author} {\bibfnamefont {A.}~\bibnamefont {Megrant}}, \bibinfo
  {author} {\bibfnamefont {J.}~\bibnamefont {Mutus}}, \bibinfo {author}
  {\bibfnamefont {M.}~\bibnamefont {Neeley}}, \bibinfo {author} {\bibfnamefont
  {C.}~\bibnamefont {Neill}}, \bibinfo {author} {\bibfnamefont {P.~J.~J.}\
  \bibnamefont {O'Malley}}, \bibinfo {author} {\bibfnamefont {P.}~\bibnamefont
  {Roushan}}, \bibinfo {author} {\bibfnamefont {D.}~\bibnamefont {Sank}},
  \bibinfo {author} {\bibfnamefont {A.}~\bibnamefont {Vainsencher}}, \bibinfo
  {author} {\bibfnamefont {J.}~\bibnamefont {Wenner}}, \bibinfo {author}
  {\bibfnamefont {T.~C.}\ \bibnamefont {White}}, \bibinfo {author}
  {\bibfnamefont {A.~N.}\ \bibnamefont {Korotkov}},\ and\ \bibinfo {author}
  {\bibfnamefont {J.~M.}\ \bibnamefont {Martinis}},\ }\href
  {https://doi.org/10.1103/PhysRevLett.116.020501} {\bibfield  {journal}
  {\bibinfo  {journal} {Phys. Rev. Lett.}\ }\textbf {\bibinfo {volume} {116}},\
  \bibinfo {pages} {020501} (\bibinfo {year} {2016})}\BibitemShut {NoStop}%
\bibitem [{\citenamefont {Haupt}\ and\ \citenamefont
  {Egger}(2023)}]{Haupt_2023}%
  \BibitemOpen
  \bibfield  {author} {\bibinfo {author} {\bibfnamefont {C.~J.}\ \bibnamefont
  {Haupt}}\ and\ \bibinfo {author} {\bibfnamefont {D.~J.}\ \bibnamefont
  {Egger}},\ }\href {https://doi.org/10.1103/PhysRevA.108.022614} {\bibfield
  {journal} {\bibinfo  {journal} {Phys. Rev. A}\ }\textbf {\bibinfo {volume}
  {108}},\ \bibinfo {pages} {022614} (\bibinfo {year} {2023})}\BibitemShut
  {NoStop}%
\bibitem [{\citenamefont {Werninghaus}\ \emph {et~al.}(2021)\citenamefont
  {Werninghaus}, \citenamefont {Egger}, \citenamefont {Roy}, \citenamefont
  {Machnes}, \citenamefont {Wilhelm},\ and\ \citenamefont
  {Filipp}}]{Werninghaus_2021}%
  \BibitemOpen
  \bibfield  {author} {\bibinfo {author} {\bibfnamefont {M.}~\bibnamefont
  {Werninghaus}}, \bibinfo {author} {\bibfnamefont {D.~J.}\ \bibnamefont
  {Egger}}, \bibinfo {author} {\bibfnamefont {F.}~\bibnamefont {Roy}}, \bibinfo
  {author} {\bibfnamefont {S.}~\bibnamefont {Machnes}}, \bibinfo {author}
  {\bibfnamefont {F.~K.}\ \bibnamefont {Wilhelm}},\ and\ \bibinfo {author}
  {\bibfnamefont {S.}~\bibnamefont {Filipp}},\ }\href
  {https://doi.org/10.1038/s41534-020-00346-2} {\bibfield  {journal} {\bibinfo
  {journal} {npj Quantum Inf.}\ }\textbf {\bibinfo {volume} {7}},\ \bibinfo
  {pages} {14} (\bibinfo {year} {2021})}\BibitemShut {NoStop}%
\bibitem [{\citenamefont {Varbanov}\ \emph {et~al.}(2020)\citenamefont
  {Varbanov}, \citenamefont {Battistel}, \citenamefont {Tarasinski},
  \citenamefont {Ostroukh}, \citenamefont {O’Brien}, \citenamefont
  {DiCarlo},\ and\ \citenamefont {Terhal}}]{Varbanov_2020}%
  \BibitemOpen
  \bibfield  {author} {\bibinfo {author} {\bibfnamefont {B.~M.}\ \bibnamefont
  {Varbanov}}, \bibinfo {author} {\bibfnamefont {F.}~\bibnamefont {Battistel}},
  \bibinfo {author} {\bibfnamefont {B.~M.}\ \bibnamefont {Tarasinski}},
  \bibinfo {author} {\bibfnamefont {V.~P.}\ \bibnamefont {Ostroukh}}, \bibinfo
  {author} {\bibfnamefont {T.~E.}\ \bibnamefont {O’Brien}}, \bibinfo {author}
  {\bibfnamefont {L.}~\bibnamefont {DiCarlo}},\ and\ \bibinfo {author}
  {\bibfnamefont {B.~M.}\ \bibnamefont {Terhal}},\ }\href
  {https://doi.org/10.1038/s41534-020-00330-w} {\bibfield  {journal} {\bibinfo
  {journal} {npj Quantum Inf.}\ }\textbf {\bibinfo {volume} {6}},\ \bibinfo
  {pages} {102} (\bibinfo {year} {2020})}\BibitemShut {NoStop}%
\bibitem [{\citenamefont {McEwen}\ \emph {et~al.}(2021)\citenamefont {McEwen},
  \citenamefont {Kafri}, \citenamefont {Chen}, \citenamefont {Atalaya},
  \citenamefont {Satzinger}, \citenamefont {Quintana}, \citenamefont {Klimov},
  \citenamefont {Sank}, \citenamefont {Gidney}, \citenamefont {Fowler},
  \citenamefont {Arute}, \citenamefont {Arya}, \citenamefont {Buckley},
  \citenamefont {Burkett}, \citenamefont {Bushnell}, \citenamefont {Chiaro},
  \citenamefont {Collins}, \citenamefont {Demura}, \citenamefont {Dunsworth},
  \citenamefont {Erickson}, \citenamefont {Foxen}, \citenamefont {Giustina},
  \citenamefont {Huang}, \citenamefont {Hong}, \citenamefont {Jeffrey},
  \citenamefont {Kim}, \citenamefont {Kechedzhi}, \citenamefont {Kostritsa},
  \citenamefont {Laptev}, \citenamefont {Megrant}, \citenamefont {Mi},
  \citenamefont {Mutus}, \citenamefont {Naaman}, \citenamefont {Neeley},
  \citenamefont {Neill}, \citenamefont {Niu}, \citenamefont {Paler},
  \citenamefont {Redd}, \citenamefont {Roushan}, \citenamefont {White},
  \citenamefont {Yao}, \citenamefont {Yeh}, \citenamefont {Zalcman},
  \citenamefont {Chen}, \citenamefont {Smelyanskiy}, \citenamefont {Martinis},
  \citenamefont {Neven}, \citenamefont {Kelly}, \citenamefont {Korotkov},
  \citenamefont {Petukhov},\ and\ \citenamefont {Barends}}]{McEwen_2021}%
  \BibitemOpen
  \bibfield  {author} {\bibinfo {author} {\bibfnamefont {M.}~\bibnamefont
  {McEwen}}, \bibinfo {author} {\bibfnamefont {D.}~\bibnamefont {Kafri}},
  \bibinfo {author} {\bibfnamefont {Z.}~\bibnamefont {Chen}}, \bibinfo {author}
  {\bibfnamefont {J.}~\bibnamefont {Atalaya}}, \bibinfo {author} {\bibfnamefont
  {K.~J.}\ \bibnamefont {Satzinger}}, \bibinfo {author} {\bibfnamefont
  {C.}~\bibnamefont {Quintana}}, \bibinfo {author} {\bibfnamefont {P.~V.}\
  \bibnamefont {Klimov}}, \bibinfo {author} {\bibfnamefont {D.}~\bibnamefont
  {Sank}}, \bibinfo {author} {\bibfnamefont {C.}~\bibnamefont {Gidney}},
  \bibinfo {author} {\bibfnamefont {A.~G.}\ \bibnamefont {Fowler}}, \bibinfo
  {author} {\bibfnamefont {F.}~\bibnamefont {Arute}}, \bibinfo {author}
  {\bibfnamefont {K.}~\bibnamefont {Arya}}, \bibinfo {author} {\bibfnamefont
  {B.}~\bibnamefont {Buckley}}, \bibinfo {author} {\bibfnamefont
  {B.}~\bibnamefont {Burkett}}, \bibinfo {author} {\bibfnamefont
  {N.}~\bibnamefont {Bushnell}}, \bibinfo {author} {\bibfnamefont
  {B.}~\bibnamefont {Chiaro}}, \bibinfo {author} {\bibfnamefont
  {R.}~\bibnamefont {Collins}}, \bibinfo {author} {\bibfnamefont
  {S.}~\bibnamefont {Demura}}, \bibinfo {author} {\bibfnamefont
  {A.}~\bibnamefont {Dunsworth}}, \bibinfo {author} {\bibfnamefont
  {C.}~\bibnamefont {Erickson}}, \bibinfo {author} {\bibfnamefont
  {B.}~\bibnamefont {Foxen}}, \bibinfo {author} {\bibfnamefont
  {M.}~\bibnamefont {Giustina}}, \bibinfo {author} {\bibfnamefont
  {T.}~\bibnamefont {Huang}}, \bibinfo {author} {\bibfnamefont
  {S.}~\bibnamefont {Hong}}, \bibinfo {author} {\bibfnamefont {E.}~\bibnamefont
  {Jeffrey}}, \bibinfo {author} {\bibfnamefont {S.}~\bibnamefont {Kim}},
  \bibinfo {author} {\bibfnamefont {K.}~\bibnamefont {Kechedzhi}}, \bibinfo
  {author} {\bibfnamefont {F.}~\bibnamefont {Kostritsa}}, \bibinfo {author}
  {\bibfnamefont {P.}~\bibnamefont {Laptev}}, \bibinfo {author} {\bibfnamefont
  {A.}~\bibnamefont {Megrant}}, \bibinfo {author} {\bibfnamefont
  {X.}~\bibnamefont {Mi}}, \bibinfo {author} {\bibfnamefont {J.}~\bibnamefont
  {Mutus}}, \bibinfo {author} {\bibfnamefont {O.}~\bibnamefont {Naaman}},
  \bibinfo {author} {\bibfnamefont {M.}~\bibnamefont {Neeley}}, \bibinfo
  {author} {\bibfnamefont {C.}~\bibnamefont {Neill}}, \bibinfo {author}
  {\bibfnamefont {M.}~\bibnamefont {Niu}}, \bibinfo {author} {\bibfnamefont
  {A.}~\bibnamefont {Paler}}, \bibinfo {author} {\bibfnamefont
  {N.}~\bibnamefont {Redd}}, \bibinfo {author} {\bibfnamefont {P.}~\bibnamefont
  {Roushan}}, \bibinfo {author} {\bibfnamefont {T.~C.}\ \bibnamefont {White}},
  \bibinfo {author} {\bibfnamefont {J.}~\bibnamefont {Yao}}, \bibinfo {author}
  {\bibfnamefont {P.}~\bibnamefont {Yeh}}, \bibinfo {author} {\bibfnamefont
  {A.}~\bibnamefont {Zalcman}}, \bibinfo {author} {\bibfnamefont
  {Y.}~\bibnamefont {Chen}}, \bibinfo {author} {\bibfnamefont {V.~N.}\
  \bibnamefont {Smelyanskiy}}, \bibinfo {author} {\bibfnamefont {J.~M.}\
  \bibnamefont {Martinis}}, \bibinfo {author} {\bibfnamefont {H.}~\bibnamefont
  {Neven}}, \bibinfo {author} {\bibfnamefont {J.}~\bibnamefont {Kelly}},
  \bibinfo {author} {\bibfnamefont {A.~N.}\ \bibnamefont {Korotkov}}, \bibinfo
  {author} {\bibfnamefont {A.~G.}\ \bibnamefont {Petukhov}},\ and\ \bibinfo
  {author} {\bibfnamefont {R.}~\bibnamefont {Barends}},\ }\href
  {https://doi.org/10.1038/s41467-021-21982-y} {\bibfield  {journal} {\bibinfo
  {journal} {Nat. Commun.}\ }\textbf {\bibinfo {volume} {12}},\ \bibinfo
  {pages} {1761} (\bibinfo {year} {2021})}\BibitemShut {NoStop}%
\bibitem [{\citenamefont {Roy}\ \emph {et~al.}(2023)\citenamefont {Roy},
  \citenamefont {Li}, \citenamefont {Kapit},\ and\ \citenamefont
  {Schuster}}]{Roy_2022}%
  \BibitemOpen
  \bibfield  {author} {\bibinfo {author} {\bibfnamefont {T.}~\bibnamefont
  {Roy}}, \bibinfo {author} {\bibfnamefont {Z.}~\bibnamefont {Li}}, \bibinfo
  {author} {\bibfnamefont {E.}~\bibnamefont {Kapit}},\ and\ \bibinfo {author}
  {\bibfnamefont {D.}~\bibnamefont {Schuster}},\ }\href
  {https://doi.org/10.1103/PhysRevApplied.19.064024} {\bibfield  {journal}
  {\bibinfo  {journal} {Phys. Rev. Appl.}\ }\textbf {\bibinfo {volume} {19}},\
  \bibinfo {pages} {064024} (\bibinfo {year} {2023})}\BibitemShut {NoStop}%
\bibitem [{\citenamefont {Gokhale}\ \emph {et~al.}(2019)\citenamefont
  {Gokhale}, \citenamefont {Baker}, \citenamefont {Duckering}, \citenamefont
  {Brown}, \citenamefont {Brown},\ and\ \citenamefont {Chong}}]{Gokhale_2019}%
  \BibitemOpen
  \bibfield  {author} {\bibinfo {author} {\bibfnamefont {P.}~\bibnamefont
  {Gokhale}}, \bibinfo {author} {\bibfnamefont {J.~M.}\ \bibnamefont {Baker}},
  \bibinfo {author} {\bibfnamefont {C.}~\bibnamefont {Duckering}}, \bibinfo
  {author} {\bibfnamefont {N.~C.}\ \bibnamefont {Brown}}, \bibinfo {author}
  {\bibfnamefont {K.~R.}\ \bibnamefont {Brown}},\ and\ \bibinfo {author}
  {\bibfnamefont {F.~T.}\ \bibnamefont {Chong}},\ }in\ \href
  {https://doi.org/10.1145/3307650.3322253} {\emph {\bibinfo {booktitle}
  {Proceedings of the 46th International Symposium on Computer
  Architecture}}},\ \bibinfo {series and number} {ISCA'19}\ (\bibinfo
  {publisher} {Association for Computing Machinery},\ \bibinfo {address} {New
  York, USA},\ \bibinfo {year} {2019})\ pp.\ \bibinfo {pages}
  {554--566}\BibitemShut {NoStop}%
\bibitem [{\citenamefont {Galda}\ \emph {et~al.}(2021)\citenamefont {Galda},
  \citenamefont {Cubeddu}, \citenamefont {Kanazawa}, \citenamefont {Narang},\
  and\ \citenamefont {Earnest-Noble}}]{Galda_2021}%
  \BibitemOpen
  \bibfield  {author} {\bibinfo {author} {\bibfnamefont {A.}~\bibnamefont
  {Galda}}, \bibinfo {author} {\bibfnamefont {M.}~\bibnamefont {Cubeddu}},
  \bibinfo {author} {\bibfnamefont {N.}~\bibnamefont {Kanazawa}}, \bibinfo
  {author} {\bibfnamefont {P.}~\bibnamefont {Narang}},\ and\ \bibinfo {author}
  {\bibfnamefont {N.}~\bibnamefont {Earnest-Noble}},\ }\href
  {https://doi.org/10.48550/arXiv.2109.00558} {\bibfield  {journal} {\bibinfo
  {journal} {arXiv.2109.00558}\ } (\bibinfo {year} {2021})}\BibitemShut
  {NoStop}%
\bibitem [{\citenamefont {Saha}\ and\ \citenamefont
  {Khanna}(2023)}]{Saha_2023}%
  \BibitemOpen
  \bibfield  {author} {\bibinfo {author} {\bibfnamefont {A.}~\bibnamefont
  {Saha}}\ and\ \bibinfo {author} {\bibfnamefont {O.}~\bibnamefont {Khanna}},\
  }\href {https://doi.org/10.48550/arXiv.2304.03050} {\bibfield  {journal}
  {\bibinfo  {journal} {arXiv.2304.03050}\ } (\bibinfo {year}
  {2023})}\BibitemShut {NoStop}%
\bibitem [{\citenamefont {Kurpiers}\ \emph {et~al.}(2018)\citenamefont
  {Kurpiers}, \citenamefont {Magnard}, \citenamefont {Walter}, \citenamefont
  {Royer}, \citenamefont {Pechal}, \citenamefont {Heinsoo}, \citenamefont
  {Salathé}, \citenamefont {Akin}, \citenamefont {Storz}, \citenamefont
  {Besse}, \citenamefont {Gasparinetti}, \citenamefont {Blais},\ and\
  \citenamefont {Wallraff}}]{Kurpiers_2018}%
  \BibitemOpen
  \bibfield  {author} {\bibinfo {author} {\bibfnamefont {P.}~\bibnamefont
  {Kurpiers}}, \bibinfo {author} {\bibfnamefont {P.}~\bibnamefont {Magnard}},
  \bibinfo {author} {\bibfnamefont {T.}~\bibnamefont {Walter}}, \bibinfo
  {author} {\bibfnamefont {B.}~\bibnamefont {Royer}}, \bibinfo {author}
  {\bibfnamefont {M.}~\bibnamefont {Pechal}}, \bibinfo {author} {\bibfnamefont
  {J.}~\bibnamefont {Heinsoo}}, \bibinfo {author} {\bibfnamefont
  {Y.}~\bibnamefont {Salathé}}, \bibinfo {author} {\bibfnamefont
  {A.}~\bibnamefont {Akin}}, \bibinfo {author} {\bibfnamefont {S.}~\bibnamefont
  {Storz}}, \bibinfo {author} {\bibfnamefont {J.-C.}\ \bibnamefont {Besse}},
  \bibinfo {author} {\bibfnamefont {S.}~\bibnamefont {Gasparinetti}}, \bibinfo
  {author} {\bibfnamefont {A.}~\bibnamefont {Blais}},\ and\ \bibinfo {author}
  {\bibfnamefont {A.}~\bibnamefont {Wallraff}},\ }\href
  {https://doi.org/10.1038/s41586-018-0195-y} {\bibfield  {journal} {\bibinfo
  {journal} {Nature}\ }\textbf {\bibinfo {volume} {558}},\ \bibinfo {pages}
  {264} (\bibinfo {year} {2018})}\BibitemShut {NoStop}%
\bibitem [{\citenamefont {Blok}\ \emph {et~al.}(2021)\citenamefont {Blok},
  \citenamefont {Ramasesh}, \citenamefont {Schuster}, \citenamefont {O'Brien},
  \citenamefont {Kreikebaum}, \citenamefont {Dahlen}, \citenamefont {Morvan},
  \citenamefont {Yoshida}, \citenamefont {Yao},\ and\ \citenamefont
  {Siddiqi}}]{Blok_2021}%
  \BibitemOpen
  \bibfield  {author} {\bibinfo {author} {\bibfnamefont {M.~S.}\ \bibnamefont
  {Blok}}, \bibinfo {author} {\bibfnamefont {V.~V.}\ \bibnamefont {Ramasesh}},
  \bibinfo {author} {\bibfnamefont {T.}~\bibnamefont {Schuster}}, \bibinfo
  {author} {\bibfnamefont {K.}~\bibnamefont {O'Brien}}, \bibinfo {author}
  {\bibfnamefont {J.~M.}\ \bibnamefont {Kreikebaum}}, \bibinfo {author}
  {\bibfnamefont {D.}~\bibnamefont {Dahlen}}, \bibinfo {author} {\bibfnamefont
  {A.}~\bibnamefont {Morvan}}, \bibinfo {author} {\bibfnamefont
  {B.}~\bibnamefont {Yoshida}}, \bibinfo {author} {\bibfnamefont {N.~Y.}\
  \bibnamefont {Yao}},\ and\ \bibinfo {author} {\bibfnamefont {I.}~\bibnamefont
  {Siddiqi}},\ }\href {https://doi.org/10.1103/PhysRevX.11.021010} {\bibfield
  {journal} {\bibinfo  {journal} {Phys. Rev. X}\ }\textbf {\bibinfo {volume}
  {11}},\ \bibinfo {pages} {021010} (\bibinfo {year} {2021})}\BibitemShut
  {NoStop}%
\bibitem [{\citenamefont {Chen}\ \emph {et~al.}(2023)\citenamefont {Chen},
  \citenamefont {Li}, \citenamefont {Lu}, \citenamefont {Warren}, \citenamefont
  {Kri\v{z}an}, \citenamefont {Kosen}, \citenamefont {Rommel}, \citenamefont
  {Ahmed}, \citenamefont {Osman}, \citenamefont {Bizn\'{a}rov\'{a}},
  \citenamefont {Fadavi~Roudsari}, \citenamefont {Lienhard}, \citenamefont
  {Caputo}, \citenamefont {Grigoras}, \citenamefont {Gr\"{o}nberg},
  \citenamefont {Govenius}, \citenamefont {Kockum}, \citenamefont {Delsing},
  \citenamefont {Bylander},\ and\ \citenamefont {Tancredi}}]{Chen_2023}%
  \BibitemOpen
  \bibfield  {author} {\bibinfo {author} {\bibfnamefont {L.}~\bibnamefont
  {Chen}}, \bibinfo {author} {\bibfnamefont {H.-X.}\ \bibnamefont {Li}},
  \bibinfo {author} {\bibfnamefont {Y.}~\bibnamefont {Lu}}, \bibinfo {author}
  {\bibfnamefont {C.~W.}\ \bibnamefont {Warren}}, \bibinfo {author}
  {\bibfnamefont {C.~J.}\ \bibnamefont {Kri\v{z}an}}, \bibinfo {author}
  {\bibfnamefont {S.}~\bibnamefont {Kosen}}, \bibinfo {author} {\bibfnamefont
  {M.}~\bibnamefont {Rommel}}, \bibinfo {author} {\bibfnamefont
  {S.}~\bibnamefont {Ahmed}}, \bibinfo {author} {\bibfnamefont
  {A.}~\bibnamefont {Osman}}, \bibinfo {author} {\bibfnamefont
  {J.}~\bibnamefont {Bizn\'{a}rov\'{a}}}, \bibinfo {author} {\bibfnamefont
  {A.}~\bibnamefont {Fadavi~Roudsari}}, \bibinfo {author} {\bibfnamefont
  {B.}~\bibnamefont {Lienhard}}, \bibinfo {author} {\bibfnamefont
  {M.}~\bibnamefont {Caputo}}, \bibinfo {author} {\bibfnamefont
  {K.}~\bibnamefont {Grigoras}}, \bibinfo {author} {\bibfnamefont
  {L.}~\bibnamefont {Gr\"{o}nberg}}, \bibinfo {author} {\bibfnamefont
  {J.}~\bibnamefont {Govenius}}, \bibinfo {author} {\bibfnamefont {A.~F.}\
  \bibnamefont {Kockum}}, \bibinfo {author} {\bibfnamefont {P.}~\bibnamefont
  {Delsing}}, \bibinfo {author} {\bibfnamefont {J.}~\bibnamefont {Bylander}},\
  and\ \bibinfo {author} {\bibfnamefont {G.}~\bibnamefont {Tancredi}},\ }\href
  {https://doi.org/10.1038/s41534-023-00689-6} {\bibfield  {journal} {\bibinfo
  {journal} {npj Quantum Inf.}\ }\textbf {\bibinfo {volume} {9}},\ \bibinfo
  {pages} {26} (\bibinfo {year} {2023})}\BibitemShut {NoStop}%
\bibitem [{\citenamefont {Kehrer}\ \emph {et~al.}(2022)\citenamefont {Kehrer},
  \citenamefont {Nadolny},\ and\ \citenamefont {Bruder}}]{Kehrer_2023}%
  \BibitemOpen
  \bibfield  {author} {\bibinfo {author} {\bibfnamefont {T.}~\bibnamefont
  {Kehrer}}, \bibinfo {author} {\bibfnamefont {T.}~\bibnamefont {Nadolny}},\
  and\ \bibinfo {author} {\bibfnamefont {C.}~\bibnamefont {Bruder}},\ }\href
  {https://doi.org/10.48550/arXiv.2307.13504} {\bibfield  {journal} {\bibinfo
  {journal} {arXiv.2307.13504}\ } (\bibinfo {year} {2022})}\BibitemShut
  {NoStop}%
\bibitem [{\citenamefont {Cervera-Lierta}\ \emph {et~al.}(2022)\citenamefont
  {Cervera-Lierta}, \citenamefont {Krenn}, \citenamefont {Aspuru-Guzik},\ and\
  \citenamefont {Galda}}]{Cervera-Lierta_2022}%
  \BibitemOpen
  \bibfield  {author} {\bibinfo {author} {\bibfnamefont {A.}~\bibnamefont
  {Cervera-Lierta}}, \bibinfo {author} {\bibfnamefont {M.}~\bibnamefont
  {Krenn}}, \bibinfo {author} {\bibfnamefont {A.}~\bibnamefont
  {Aspuru-Guzik}},\ and\ \bibinfo {author} {\bibfnamefont {A.}~\bibnamefont
  {Galda}},\ }\href {https://doi.org/10.1103/PhysRevApplied.17.024062}
  {\bibfield  {journal} {\bibinfo  {journal} {Phys. Rev. Appl.}\ }\textbf
  {\bibinfo {volume} {17}},\ \bibinfo {pages} {024062} (\bibinfo {year}
  {2022})}\BibitemShut {NoStop}%
\bibitem [{\citenamefont {Andrews}\ \emph {et~al.}(2019)\citenamefont
  {Andrews}, \citenamefont {Jones}, \citenamefont {Reed}, \citenamefont
  {Jones}, \citenamefont {Ha}, \citenamefont {Jura}, \citenamefont {Kerckhoff},
  \citenamefont {Levendorf}, \citenamefont {Meenehan}, \citenamefont {Merkel},
  \citenamefont {Smith}, \citenamefont {Sun}, \citenamefont {Weinstein},
  \citenamefont {Rakher}, \citenamefont {Ladd},\ and\ \citenamefont
  {Borselli}}]{Andrews_2019}%
  \BibitemOpen
  \bibfield  {author} {\bibinfo {author} {\bibfnamefont {R.~W.}\ \bibnamefont
  {Andrews}}, \bibinfo {author} {\bibfnamefont {C.}~\bibnamefont {Jones}},
  \bibinfo {author} {\bibfnamefont {M.~D.}\ \bibnamefont {Reed}}, \bibinfo
  {author} {\bibfnamefont {A.~M.}\ \bibnamefont {Jones}}, \bibinfo {author}
  {\bibfnamefont {S.~D.}\ \bibnamefont {Ha}}, \bibinfo {author} {\bibfnamefont
  {M.~P.}\ \bibnamefont {Jura}}, \bibinfo {author} {\bibfnamefont
  {J.}~\bibnamefont {Kerckhoff}}, \bibinfo {author} {\bibfnamefont
  {M.}~\bibnamefont {Levendorf}}, \bibinfo {author} {\bibfnamefont
  {S.}~\bibnamefont {Meenehan}}, \bibinfo {author} {\bibfnamefont {S.~T.}\
  \bibnamefont {Merkel}}, \bibinfo {author} {\bibfnamefont {A.}~\bibnamefont
  {Smith}}, \bibinfo {author} {\bibfnamefont {B.}~\bibnamefont {Sun}}, \bibinfo
  {author} {\bibfnamefont {A.~J.}\ \bibnamefont {Weinstein}}, \bibinfo {author}
  {\bibfnamefont {M.~T.}\ \bibnamefont {Rakher}}, \bibinfo {author}
  {\bibfnamefont {T.~D.}\ \bibnamefont {Ladd}},\ and\ \bibinfo {author}
  {\bibfnamefont {M.~G.}\ \bibnamefont {Borselli}},\ }\href
  {https://doi.org/10.1038/s41565-019-0500-4} {\bibfield  {journal} {\bibinfo
  {journal} {Nat. Nanotechnol.}\ }\textbf {\bibinfo {volume} {14}},\ \bibinfo
  {pages} {747} (\bibinfo {year} {2019})}\BibitemShut {NoStop}%
\bibitem [{\citenamefont {Deist}\ \emph {et~al.}(2022)\citenamefont {Deist},
  \citenamefont {Lu}, \citenamefont {Ho}, \citenamefont {Pasha}, \citenamefont
  {Zeiher}, \citenamefont {Yan},\ and\ \citenamefont
  {Stamper-Kurn}}]{Graham_2023}%
  \BibitemOpen
  \bibfield  {author} {\bibinfo {author} {\bibfnamefont {E.}~\bibnamefont
  {Deist}}, \bibinfo {author} {\bibfnamefont {Y.-H.}\ \bibnamefont {Lu}},
  \bibinfo {author} {\bibfnamefont {J.}~\bibnamefont {Ho}}, \bibinfo {author}
  {\bibfnamefont {M.~K.}\ \bibnamefont {Pasha}}, \bibinfo {author}
  {\bibfnamefont {J.}~\bibnamefont {Zeiher}}, \bibinfo {author} {\bibfnamefont
  {Z.}~\bibnamefont {Yan}},\ and\ \bibinfo {author} {\bibfnamefont {D.~M.}\
  \bibnamefont {Stamper-Kurn}},\ }\href
  {https://doi.org/10.1103/PhysRevLett.129.203602} {\bibfield  {journal}
  {\bibinfo  {journal} {Phys. Rev. Lett.}\ }\textbf {\bibinfo {volume} {129}},\
  \bibinfo {pages} {203602} (\bibinfo {year} {2022})}\BibitemShut {NoStop}%
\bibitem [{\citenamefont {Govia}\ \emph {et~al.}(2022)\citenamefont {Govia},
  \citenamefont {Jurcevic}, \citenamefont {Merkel},\ and\ \citenamefont
  {McKay}}]{Govia_2022}%
  \BibitemOpen
  \bibfield  {author} {\bibinfo {author} {\bibfnamefont {L.~C.~G.}\
  \bibnamefont {Govia}}, \bibinfo {author} {\bibfnamefont {P.}~\bibnamefont
  {Jurcevic}}, \bibinfo {author} {\bibfnamefont {S.~T.}\ \bibnamefont
  {Merkel}},\ and\ \bibinfo {author} {\bibfnamefont {D.~C.}\ \bibnamefont
  {McKay}},\ }\href {https://doi.org/10.48550/arXiv.2207.04836} {\bibfield
  {journal} {\bibinfo  {journal} {arXiv.2207.04836}\ } (\bibinfo {year}
  {2022})}\BibitemShut {NoStop}%
\bibitem [{\citenamefont {Kanazawa}\ \emph {et~al.}(2023)\citenamefont
  {Kanazawa}, \citenamefont {Egger}, \citenamefont {Ben-Haim}, \citenamefont
  {Zhang}, \citenamefont {Shanks}, \citenamefont {Aleksandrowicz},\ and\
  \citenamefont {Wood}}]{Kanazawa_2023}%
  \BibitemOpen
  \bibfield  {author} {\bibinfo {author} {\bibfnamefont {N.}~\bibnamefont
  {Kanazawa}}, \bibinfo {author} {\bibfnamefont {D.~J.}\ \bibnamefont {Egger}},
  \bibinfo {author} {\bibfnamefont {Y.}~\bibnamefont {Ben-Haim}}, \bibinfo
  {author} {\bibfnamefont {H.}~\bibnamefont {Zhang}}, \bibinfo {author}
  {\bibfnamefont {W.~E.}\ \bibnamefont {Shanks}}, \bibinfo {author}
  {\bibfnamefont {G.}~\bibnamefont {Aleksandrowicz}},\ and\ \bibinfo {author}
  {\bibfnamefont {C.~L.}\ \bibnamefont {Wood}},\ }\href
  {https://doi.org/10.21105/joss.05329} {\bibfield  {journal} {\bibinfo
  {journal} {J. Open Source Softw.}\ }\textbf {\bibinfo {volume} {8}},\
  \bibinfo {pages} {5329} (\bibinfo {year} {2023})}\BibitemShut {NoStop}%
\bibitem [{\citenamefont {Sank}\ \emph {et~al.}(2016)\citenamefont {Sank},
  \citenamefont {Chen}, \citenamefont {Khezri}, \citenamefont {Kelly},
  \citenamefont {Barends}, \citenamefont {Campbell}, \citenamefont {Chen},
  \citenamefont {Chiaro}, \citenamefont {Dunsworth}, \citenamefont {Fowler},
  \citenamefont {Jeffrey}, \citenamefont {Lucero}, \citenamefont {Megrant},
  \citenamefont {Mutus}, \citenamefont {Neeley}, \citenamefont {Neill},
  \citenamefont {O'Malley}, \citenamefont {Quintana}, \citenamefont {Roushan},
  \citenamefont {Vainsencher}, \citenamefont {White}, \citenamefont {Wenner},
  \citenamefont {Korotkov},\ and\ \citenamefont {Martinis}}]{Sank_2016}%
  \BibitemOpen
  \bibfield  {author} {\bibinfo {author} {\bibfnamefont {D.}~\bibnamefont
  {Sank}}, \bibinfo {author} {\bibfnamefont {Z.}~\bibnamefont {Chen}}, \bibinfo
  {author} {\bibfnamefont {M.}~\bibnamefont {Khezri}}, \bibinfo {author}
  {\bibfnamefont {J.}~\bibnamefont {Kelly}}, \bibinfo {author} {\bibfnamefont
  {R.}~\bibnamefont {Barends}}, \bibinfo {author} {\bibfnamefont
  {B.}~\bibnamefont {Campbell}}, \bibinfo {author} {\bibfnamefont
  {Y.}~\bibnamefont {Chen}}, \bibinfo {author} {\bibfnamefont {B.}~\bibnamefont
  {Chiaro}}, \bibinfo {author} {\bibfnamefont {A.}~\bibnamefont {Dunsworth}},
  \bibinfo {author} {\bibfnamefont {A.}~\bibnamefont {Fowler}}, \bibinfo
  {author} {\bibfnamefont {E.}~\bibnamefont {Jeffrey}}, \bibinfo {author}
  {\bibfnamefont {E.}~\bibnamefont {Lucero}}, \bibinfo {author} {\bibfnamefont
  {A.}~\bibnamefont {Megrant}}, \bibinfo {author} {\bibfnamefont
  {J.}~\bibnamefont {Mutus}}, \bibinfo {author} {\bibfnamefont
  {M.}~\bibnamefont {Neeley}}, \bibinfo {author} {\bibfnamefont
  {C.}~\bibnamefont {Neill}}, \bibinfo {author} {\bibfnamefont {P.~J.~J.}\
  \bibnamefont {O'Malley}}, \bibinfo {author} {\bibfnamefont {C.}~\bibnamefont
  {Quintana}}, \bibinfo {author} {\bibfnamefont {P.}~\bibnamefont {Roushan}},
  \bibinfo {author} {\bibfnamefont {A.}~\bibnamefont {Vainsencher}}, \bibinfo
  {author} {\bibfnamefont {T.}~\bibnamefont {White}}, \bibinfo {author}
  {\bibfnamefont {J.}~\bibnamefont {Wenner}}, \bibinfo {author} {\bibfnamefont
  {A.~N.}\ \bibnamefont {Korotkov}},\ and\ \bibinfo {author} {\bibfnamefont
  {J.~M.}\ \bibnamefont {Martinis}},\ }\href
  {https://doi.org/10.1103/PhysRevLett.117.190503} {\bibfield  {journal}
  {\bibinfo  {journal} {Phys. Rev. Lett.}\ }\textbf {\bibinfo {volume} {117}},\
  \bibinfo {pages} {190503} (\bibinfo {year} {2016})}\BibitemShut {NoStop}%
\bibitem [{\citenamefont {Maciejewski}\ \emph {et~al.}(2020)\citenamefont
  {Maciejewski}, \citenamefont {Zimbor{\'{a}}s},\ and\ \citenamefont
  {Oszmaniec}}]{Maciejewski_2020}%
  \BibitemOpen
  \bibfield  {author} {\bibinfo {author} {\bibfnamefont {F.~B.}\ \bibnamefont
  {Maciejewski}}, \bibinfo {author} {\bibfnamefont {Z.}~\bibnamefont
  {Zimbor{\'{a}}s}},\ and\ \bibinfo {author} {\bibfnamefont {M.}~\bibnamefont
  {Oszmaniec}},\ }\href {https://doi.org/10.22331/q-2020-04-24-257} {\bibfield
  {journal} {\bibinfo  {journal} {Quantum}\ }\textbf {\bibinfo {volume} {4}},\
  \bibinfo {pages} {257} (\bibinfo {year} {2020})}\BibitemShut {NoStop}%
\bibitem [{\citenamefont {Pedregosa}\ \emph {et~al.}(2011)\citenamefont
  {Pedregosa}, \citenamefont {Varoquaux}, \citenamefont {Gramfort},
  \citenamefont {Michel}, \citenamefont {Thirion}, \citenamefont {Grisel},
  \citenamefont {Blondel}, \citenamefont {Prettenhofer}, \citenamefont {Weiss},
  \citenamefont {Dubourg}, \citenamefont {Vanderplas}, \citenamefont {Passos},
  \citenamefont {Cournapeau}, \citenamefont {Brucher}, \citenamefont {Perrot},\
  and\ \citenamefont {Duchesnay}}]{scikit-learn}%
  \BibitemOpen
  \bibfield  {author} {\bibinfo {author} {\bibfnamefont {F.}~\bibnamefont
  {Pedregosa}}, \bibinfo {author} {\bibfnamefont {G.}~\bibnamefont
  {Varoquaux}}, \bibinfo {author} {\bibfnamefont {A.}~\bibnamefont {Gramfort}},
  \bibinfo {author} {\bibfnamefont {V.}~\bibnamefont {Michel}}, \bibinfo
  {author} {\bibfnamefont {B.}~\bibnamefont {Thirion}}, \bibinfo {author}
  {\bibfnamefont {O.}~\bibnamefont {Grisel}}, \bibinfo {author} {\bibfnamefont
  {M.}~\bibnamefont {Blondel}}, \bibinfo {author} {\bibfnamefont
  {P.}~\bibnamefont {Prettenhofer}}, \bibinfo {author} {\bibfnamefont
  {R.}~\bibnamefont {Weiss}}, \bibinfo {author} {\bibfnamefont
  {V.}~\bibnamefont {Dubourg}}, \bibinfo {author} {\bibfnamefont
  {J.}~\bibnamefont {Vanderplas}}, \bibinfo {author} {\bibfnamefont
  {A.}~\bibnamefont {Passos}}, \bibinfo {author} {\bibfnamefont
  {D.}~\bibnamefont {Cournapeau}}, \bibinfo {author} {\bibfnamefont
  {M.}~\bibnamefont {Brucher}}, \bibinfo {author} {\bibfnamefont
  {M.}~\bibnamefont {Perrot}},\ and\ \bibinfo {author} {\bibfnamefont
  {{\'E}.}~\bibnamefont {Duchesnay}},\ }\href
  {http://jmlr.org/papers/v12/pedregosa11a.html} {\bibfield  {journal}
  {\bibinfo  {journal} {J. Mach. Learn. Res.}\ }\textbf {\bibinfo {volume}
  {12}},\ \bibinfo {pages} {2825} (\bibinfo {year} {2011})}\BibitemShut
  {NoStop}%
\bibitem [{\citenamefont {McKay}\ \emph {et~al.}(2017)\citenamefont {McKay},
  \citenamefont {Wood}, \citenamefont {Sheldon}, \citenamefont {Chow},\ and\
  \citenamefont {Gambetta}}]{McKay_2017}%
  \BibitemOpen
  \bibfield  {author} {\bibinfo {author} {\bibfnamefont {D.~C.}\ \bibnamefont
  {McKay}}, \bibinfo {author} {\bibfnamefont {C.~J.}\ \bibnamefont {Wood}},
  \bibinfo {author} {\bibfnamefont {S.}~\bibnamefont {Sheldon}}, \bibinfo
  {author} {\bibfnamefont {J.~M.}\ \bibnamefont {Chow}},\ and\ \bibinfo
  {author} {\bibfnamefont {J.~M.}\ \bibnamefont {Gambetta}},\ }\href
  {https://doi.org/10.1103/PhysRevA.96.022330} {\bibfield  {journal} {\bibinfo
  {journal} {Phys. Rev. A}\ }\textbf {\bibinfo {volume} {96}},\ \bibinfo
  {pages} {022330} (\bibinfo {year} {2017})}\BibitemShut {NoStop}%
\bibitem [{\citenamefont {Smolin}\ \emph {et~al.}(2012)\citenamefont {Smolin},
  \citenamefont {Gambetta},\ and\ \citenamefont {Smith}}]{Smolin_2012}%
  \BibitemOpen
  \bibfield  {author} {\bibinfo {author} {\bibfnamefont {J.~A.}\ \bibnamefont
  {Smolin}}, \bibinfo {author} {\bibfnamefont {J.~M.}\ \bibnamefont
  {Gambetta}},\ and\ \bibinfo {author} {\bibfnamefont {G.}~\bibnamefont
  {Smith}},\ }\href {https://doi.org/10.1103/PhysRevLett.108.070502} {\bibfield
   {journal} {\bibinfo  {journal} {Phys. Rev. Lett.}\ }\textbf {\bibinfo
  {volume} {108}},\ \bibinfo {pages} {070502} (\bibinfo {year}
  {2012})}\BibitemShut {NoStop}%
\bibitem [{\citenamefont {McClure}\ \emph {et~al.}(2016)\citenamefont
  {McClure}, \citenamefont {Paik}, \citenamefont {Bishop}, \citenamefont
  {Steffen}, \citenamefont {Chow},\ and\ \citenamefont
  {Gambetta}}]{McClure_2016}%
  \BibitemOpen
  \bibfield  {author} {\bibinfo {author} {\bibfnamefont {D.}~\bibnamefont
  {McClure}}, \bibinfo {author} {\bibfnamefont {H.}~\bibnamefont {Paik}},
  \bibinfo {author} {\bibfnamefont {L.}~\bibnamefont {Bishop}}, \bibinfo
  {author} {\bibfnamefont {M.}~\bibnamefont {Steffen}}, \bibinfo {author}
  {\bibfnamefont {J.~M.}\ \bibnamefont {Chow}},\ and\ \bibinfo {author}
  {\bibfnamefont {J.~M.}\ \bibnamefont {Gambetta}},\ }\bibfield  {journal}
  {\bibinfo  {journal} {Physical Review Applied}\ }\textbf {\bibinfo {volume}
  {5}},\ \href {https://doi.org/10.1103/physrevapplied.5.011001}
  {10.1103/physrevapplied.5.011001} (\bibinfo {year} {2016})\BibitemShut
  {NoStop}%
\bibitem [{\citenamefont {Rudinger}\ \emph {et~al.}(2021)\citenamefont
  {Rudinger}, \citenamefont {Ribeill}, \citenamefont {Govia}, \citenamefont
  {Ware}, \citenamefont {Nielsen}, \citenamefont {Young}, \citenamefont {Ohki},
  \citenamefont {Blume-Kohout},\ and\ \citenamefont
  {Proctor}}]{rudinger2021characterizing}%
  \BibitemOpen
  \bibfield  {author} {\bibinfo {author} {\bibfnamefont {K.}~\bibnamefont
  {Rudinger}}, \bibinfo {author} {\bibfnamefont {G.~J.}\ \bibnamefont
  {Ribeill}}, \bibinfo {author} {\bibfnamefont {L.~C.~G.}\ \bibnamefont
  {Govia}}, \bibinfo {author} {\bibfnamefont {M.}~\bibnamefont {Ware}},
  \bibinfo {author} {\bibfnamefont {E.}~\bibnamefont {Nielsen}}, \bibinfo
  {author} {\bibfnamefont {K.}~\bibnamefont {Young}}, \bibinfo {author}
  {\bibfnamefont {T.~A.}\ \bibnamefont {Ohki}}, \bibinfo {author}
  {\bibfnamefont {R.}~\bibnamefont {Blume-Kohout}},\ and\ \bibinfo {author}
  {\bibfnamefont {T.}~\bibnamefont {Proctor}},\ }\href@noop {} {\bibinfo
  {title} {Characterizing mid-circuit measurements on a superconducting qubit
  using gate set tomography}} (\bibinfo {year} {2021}),\ \Eprint
  {https://arxiv.org/abs/2103.03008} {arXiv:2103.03008 [quant-ph]} \BibitemShut
  {NoStop}%
\bibitem [{\citenamefont {Wood}\ and\ \citenamefont
  {Gambetta}(2018)}]{Wood_2018}%
  \BibitemOpen
  \bibfield  {author} {\bibinfo {author} {\bibfnamefont {C.~J.}\ \bibnamefont
  {Wood}}\ and\ \bibinfo {author} {\bibfnamefont {J.~M.}\ \bibnamefont
  {Gambetta}},\ }\href {https://doi.org/10.1103/PhysRevA.97.032306} {\bibfield
  {journal} {\bibinfo  {journal} {Phys. Rev. A}\ }\textbf {\bibinfo {volume}
  {97}},\ \bibinfo {pages} {032306} (\bibinfo {year} {2018})}\BibitemShut
  {NoStop}%
\bibitem [{\citenamefont {Peterer}\ \emph {et~al.}(2015)\citenamefont
  {Peterer}, \citenamefont {Bader}, \citenamefont {Jin}, \citenamefont {Yan},
  \citenamefont {Kamal}, \citenamefont {Gudmundsen}, \citenamefont {Leek},
  \citenamefont {Orlando}, \citenamefont {Oliver},\ and\ \citenamefont
  {Gustavsson}}]{Peterer_2015}%
  \BibitemOpen
  \bibfield  {author} {\bibinfo {author} {\bibfnamefont {M.~J.}\ \bibnamefont
  {Peterer}}, \bibinfo {author} {\bibfnamefont {S.~J.}\ \bibnamefont {Bader}},
  \bibinfo {author} {\bibfnamefont {X.}~\bibnamefont {Jin}}, \bibinfo {author}
  {\bibfnamefont {F.}~\bibnamefont {Yan}}, \bibinfo {author} {\bibfnamefont
  {A.}~\bibnamefont {Kamal}}, \bibinfo {author} {\bibfnamefont {T.~J.}\
  \bibnamefont {Gudmundsen}}, \bibinfo {author} {\bibfnamefont {P.~J.}\
  \bibnamefont {Leek}}, \bibinfo {author} {\bibfnamefont {T.~P.}\ \bibnamefont
  {Orlando}}, \bibinfo {author} {\bibfnamefont {W.~D.}\ \bibnamefont
  {Oliver}},\ and\ \bibinfo {author} {\bibfnamefont {S.}~\bibnamefont
  {Gustavsson}},\ }\href {https://doi.org/10.1103/PhysRevLett.114.010501}
  {\bibfield  {journal} {\bibinfo  {journal} {Phys. Rev. Lett.}\ }\textbf
  {\bibinfo {volume} {114}},\ \bibinfo {pages} {010501} (\bibinfo {year}
  {2015})}\BibitemShut {NoStop}%
\bibitem [{\citenamefont {Bronn}\ \emph {et~al.}(2017)\citenamefont {Bronn},
  \citenamefont {Abdo}, \citenamefont {Inoue}, \citenamefont {Lekuch},
  \citenamefont {C\'{o}rcoles}, \citenamefont {Hertzberg}, \citenamefont
  {Takita}, \citenamefont {Bishop}, \citenamefont {Gambetta},\ and\
  \citenamefont {Chow}}]{Bronn_2017}%
  \BibitemOpen
  \bibfield  {author} {\bibinfo {author} {\bibfnamefont {N.~T.}\ \bibnamefont
  {Bronn}}, \bibinfo {author} {\bibfnamefont {B.}~\bibnamefont {Abdo}},
  \bibinfo {author} {\bibfnamefont {K.}~\bibnamefont {Inoue}}, \bibinfo
  {author} {\bibfnamefont {S.}~\bibnamefont {Lekuch}}, \bibinfo {author}
  {\bibfnamefont {A.~D.}\ \bibnamefont {C\'{o}rcoles}}, \bibinfo {author}
  {\bibfnamefont {J.~B.}\ \bibnamefont {Hertzberg}}, \bibinfo {author}
  {\bibfnamefont {M.}~\bibnamefont {Takita}}, \bibinfo {author} {\bibfnamefont
  {L.~S.}\ \bibnamefont {Bishop}}, \bibinfo {author} {\bibfnamefont {J.~M.}\
  \bibnamefont {Gambetta}},\ and\ \bibinfo {author} {\bibfnamefont {J.~M.}\
  \bibnamefont {Chow}},\ }\href
  {https://doi.org/10.1088/1742-6596/834/1/012003} {\bibfield  {journal}
  {\bibinfo  {journal} {J. Phys.: Conf. Ser.}\ }\textbf {\bibinfo {volume}
  {834}},\ \bibinfo {pages} {012003} (\bibinfo {year} {2017})}\BibitemShut
  {NoStop}%
\bibitem [{\citenamefont {Fu}\ \emph {et~al.}(2018)\citenamefont {Fu},
  \citenamefont {Rol}, \citenamefont {Bultink}, \citenamefont {van Someren},
  \citenamefont {Khammassi}, \citenamefont {Ashraf}, \citenamefont {Vermeulen},
  \citenamefont {de~Sterke}, \citenamefont {Vlothuizen}, \citenamefont
  {Schouten}, \citenamefont {Almudever}, \citenamefont {DiCarlo},\ and\
  \citenamefont {Bertels}}]{Fu_2018}%
  \BibitemOpen
  \bibfield  {author} {\bibinfo {author} {\bibfnamefont {X.}~\bibnamefont
  {Fu}}, \bibinfo {author} {\bibfnamefont {M.~A.}\ \bibnamefont {Rol}},
  \bibinfo {author} {\bibfnamefont {C.~C.}\ \bibnamefont {Bultink}}, \bibinfo
  {author} {\bibfnamefont {J.}~\bibnamefont {van Someren}}, \bibinfo {author}
  {\bibfnamefont {N.}~\bibnamefont {Khammassi}}, \bibinfo {author}
  {\bibfnamefont {I.}~\bibnamefont {Ashraf}}, \bibinfo {author} {\bibfnamefont
  {R.~F.~L.}\ \bibnamefont {Vermeulen}}, \bibinfo {author} {\bibfnamefont
  {J.~C.}\ \bibnamefont {de~Sterke}}, \bibinfo {author} {\bibfnamefont {W.~J.}\
  \bibnamefont {Vlothuizen}}, \bibinfo {author} {\bibfnamefont {R.~N.}\
  \bibnamefont {Schouten}}, \bibinfo {author} {\bibfnamefont {C.~G.}\
  \bibnamefont {Almudever}}, \bibinfo {author} {\bibfnamefont {L.}~\bibnamefont
  {DiCarlo}},\ and\ \bibinfo {author} {\bibfnamefont {K.}~\bibnamefont
  {Bertels}},\ }\href {https://doi.org/10.1109/MM.2018.032271060} {\bibfield
  {journal} {\bibinfo  {journal} {IEEE Micro}\ }\textbf {\bibinfo {volume}
  {38}},\ \bibinfo {pages} {40} (\bibinfo {year} {2018})}\BibitemShut {NoStop}%
\bibitem [{\citenamefont {Underwood}\ \emph {et~al.}(2023)\citenamefont
  {Underwood}, \citenamefont {Glick}, \citenamefont {Inoue}, \citenamefont
  {Frank}, \citenamefont {Timmerwilke}, \citenamefont {Pritchett},
  \citenamefont {Chakraborty}, \citenamefont {Tien}, \citenamefont {Yeck},
  \citenamefont {Bulzacchelli}, \citenamefont {Baks}, \citenamefont {Rosno},
  \citenamefont {Robertazzi}, \citenamefont {Beck}, \citenamefont {Joshi},
  \citenamefont {Wisnieff}, \citenamefont {Ramirez}, \citenamefont {Ruedinger},
  \citenamefont {Lekuch}, \citenamefont {Gaucher},\ and\ \citenamefont
  {Friedman}}]{Underwood_2023}%
  \BibitemOpen
  \bibfield  {author} {\bibinfo {author} {\bibfnamefont {D.~L.}\ \bibnamefont
  {Underwood}}, \bibinfo {author} {\bibfnamefont {J.~A.}\ \bibnamefont
  {Glick}}, \bibinfo {author} {\bibfnamefont {K.}~\bibnamefont {Inoue}},
  \bibinfo {author} {\bibfnamefont {D.~J.}\ \bibnamefont {Frank}}, \bibinfo
  {author} {\bibfnamefont {J.}~\bibnamefont {Timmerwilke}}, \bibinfo {author}
  {\bibfnamefont {E.}~\bibnamefont {Pritchett}}, \bibinfo {author}
  {\bibfnamefont {S.}~\bibnamefont {Chakraborty}}, \bibinfo {author}
  {\bibfnamefont {K.}~\bibnamefont {Tien}}, \bibinfo {author} {\bibfnamefont
  {M.}~\bibnamefont {Yeck}}, \bibinfo {author} {\bibfnamefont {J.~F.}\
  \bibnamefont {Bulzacchelli}}, \bibinfo {author} {\bibfnamefont
  {C.}~\bibnamefont {Baks}}, \bibinfo {author} {\bibfnamefont {P.}~\bibnamefont
  {Rosno}}, \bibinfo {author} {\bibfnamefont {R.}~\bibnamefont {Robertazzi}},
  \bibinfo {author} {\bibfnamefont {M.}~\bibnamefont {Beck}}, \bibinfo {author}
  {\bibfnamefont {R.~V.}\ \bibnamefont {Joshi}}, \bibinfo {author}
  {\bibfnamefont {D.}~\bibnamefont {Wisnieff}}, \bibinfo {author}
  {\bibfnamefont {D.}~\bibnamefont {Ramirez}}, \bibinfo {author} {\bibfnamefont
  {J.}~\bibnamefont {Ruedinger}}, \bibinfo {author} {\bibfnamefont
  {S.}~\bibnamefont {Lekuch}}, \bibinfo {author} {\bibfnamefont {B.~P.}\
  \bibnamefont {Gaucher}},\ and\ \bibinfo {author} {\bibfnamefont {D.~J.}\
  \bibnamefont {Friedman}},\ }\href {https://doi.org/10.48550/arXiv.2302.11538}
  {\bibfield  {journal} {\bibinfo  {journal} {arXiv.2302.11538}\ } (\bibinfo
  {year} {2023})}\BibitemShut {NoStop}%
\bibitem [{\citenamefont {Sheldon}\ \emph {et~al.}(2016)\citenamefont
  {Sheldon}, \citenamefont {Bishop}, \citenamefont {Magesan}, \citenamefont
  {Filipp}, \citenamefont {Chow},\ and\ \citenamefont
  {Gambetta}}]{Sheldon_2016}%
  \BibitemOpen
  \bibfield  {author} {\bibinfo {author} {\bibfnamefont {S.}~\bibnamefont
  {Sheldon}}, \bibinfo {author} {\bibfnamefont {L.~S.}\ \bibnamefont {Bishop}},
  \bibinfo {author} {\bibfnamefont {E.}~\bibnamefont {Magesan}}, \bibinfo
  {author} {\bibfnamefont {S.}~\bibnamefont {Filipp}}, \bibinfo {author}
  {\bibfnamefont {J.~M.}\ \bibnamefont {Chow}},\ and\ \bibinfo {author}
  {\bibfnamefont {J.~M.}\ \bibnamefont {Gambetta}},\ }\href
  {https://doi.org/10.1103/PhysRevA.93.012301} {\bibfield  {journal} {\bibinfo
  {journal} {Phys. Rev. A}\ }\textbf {\bibinfo {volume} {93}},\ \bibinfo
  {pages} {012301} (\bibinfo {year} {2016})}\BibitemShut {NoStop}%
\end{thebibliography}%

\end{document}